\documentclass[12pt, draftclsnofoot, onecolumn]{IEEEtran}
\pdfoutput=1
\ifCLASSINFOpdf
\usepackage{graphicx}
  \graphicspath{{../pdf/}{../jpeg/}}
\DeclareGraphicsExtensions{.pdf,.jpeg,.png,.eps}
\else
\fi
\usepackage[cmex10]{amsmath}
\usepackage{amsmath,bm}
\usepackage{amssymb}
\usepackage{mathtools}

\usepackage{algorithmic}
\usepackage{algorithm}
\newsavebox{\ieeealgbox}

\usepackage[tight,footnotesize]{subfigure}
\usepackage[acronym]{glossaries}
\newacronym{mimo}{MIMO}{multiple-input multiple-output}
\newacronym{v2v}{V2V}{Vehicle-to-vehicle}
\newacronym{gscm}{GSCM}{geometric-based stochastic channel model}
\newacronym{dmc}{DMC}{dense multipath component}
\newacronym{sp}{SP}{specular paths}
\newacronym{ekf}{EKF}{extended Kalman filter}
\newacronym{sage}{SAGE}{space-alternating generalized expectation-maximization}
\newacronym{io}{IO}{interaction objects}
\newacronym{mle}{MLE}{maximum likelihood estimator}
\newacronym{blue}{BLUE}{best linear unbiased estimator}
\newacronym{tx}{TX}{transmitter}
\newacronym{rx}{RX}{receiver}
\newacronym{ddtf}{DDTF}{double directional transfer function}
\newacronym{apdp}{APDP}{average power delay profile}
\newacronym{dod}{DoD}{direction of departure}
\newacronym{doa}{DoA}{direction of arrival}
\newacronym{3d}{3D}{3-dimensional}
\newacronym{4d}{4D}{4-dimensional}
\newacronym{cdf}{CDF}{cumulative distribution function}
\newacronym{hrpe}{HRPE}{High Resolution Parameter Estimation}
\newacronym{mpc}{MPC}{multipath component}
\newacronym{fim}{FIM}{Fisher Information Matrix}
\newacronym{em}{EM}{Expectation Maximization}
\newacronym{tdm}{TDM}{time-division multiplexed}
\newacronym{snr}{SNR}{signal-to-noise ratio}
\newacronym{pdp}{PDP}{power delay profile}
\newacronym{crlb}{CRLB}{Cramer-Rao lower bound}
\newacronym{eadf}{EADF}{effective aperture distribution function}
\newacronym{rmse}{RMSE}{root mean squared error}
\newacronym{mmwave}{mmWave}{millimeter wave}
\newacronym{cmwave}{cmWave}{centimeter wave}
\newacronym{nlos}{NLOS}{non-line-of-sight}
\newacronym{ula}{ULA}{uniform linear array}
\newacronym{ura}{URA}{uniform rectangular array}
\newacronym{lp}{LP}{linear programming}
\newacronym{pd}{PD}{positive definite}
\newacronym{dut}{DUT}{device under test}
\newacronym{vna}{VNA}{vector network analyzer}
\newacronym{svd}{SVD}{singular value decomposition}
\newacronym{lra}{LRA}{low rank approximation}
\newacronym{xpd}{XPD}{cross polarization discrimination}
\newacronym{if}{IF}{intermediate frequency}
\newacronym{vmc}{VMC}{vector mixer calibration}
\newacronym{lo}{LO}{local oscillator} 
\newacronym{los}{LOS}{line-of-sight}
\newacronym{hpbw}{HPBW}{half power beamwidth}
\newacronym{fpga}{FPGA}{field-programmable gate array}
\newacronym{aps}{APS}{angular power spectrum}
\newacronym{pcb}{PCB}{printed circuit board}
\newacronym{aut}{AUT}{antenna under test}
\newacronym{pn}{PN}{phase noise}
\newacronym{pa}{PA}{power amplifier}
\newacronym{dft}{DFT}{discrete Fourier transform}
\newacronym{2d}{2D}{two-dimensional}

\newcommand{\RNum}[1]{\uppercase\expandafter{\romannumeral #1\relax}}
\newcommand{\tx}[1]{\text{#1}}
\newcommand{\B}[1]{\textbf{#1}}
\newcommand{\BS}[1]{\boldsymbol{#1}}
\newcommand\wid{0.5}

\newcommand{\ie}{\emph{i.e.}}
\newcommand{\etal}{\emph{et al. }}

\usepackage[utf8]{inputenc}
\usepackage{color}

\usepackage{amsthm}
\theoremstyle{definition}

\usepackage{siunitx}
\usepackage{booktabs}

\begin{document}
%
\title{Enabling Super-resolution Parameter Estimation for Mm-wave Channel Sounding}

\author{\IEEEauthorblockN{Rui Wang, \textit{Student Member, IEEE}, C. Umit Bas, \textit{Student Member, IEEE}, \\Zihang Cheng, \textit{Student Member, IEEE}, Thomas Choi, \textit{Student Member, IEEE}, \\ Hao Feng, \textit{Student Member, IEEE}, Zheda Li, \textit{Student Member, IEEE}, \\ Xiaokang Ye, \textit{Student Member, IEEE}, Pan Tang, \textit{Student Member, IEEE}, \\ Seun Sangodoyin, \textit{Student Member, IEEE}, Jorge G. Ponce, \textit{Student Member, IEEE}, Robert Monroe, Thomas Henige, Gary Xu, Jianzhong (Charlie) Zhang, \textit{Fellow, IEEE}, Jeongho Park, \text{Member, IEEE}, Andreas F. Molisch, \textit{Fellow, IEEE}}}


%


\maketitle

\begin{abstract}
This paper investigates the capability of millimeter-wave (mmWave) channel sounders with phased arrays to perform super-resolution parameter estimation, i.e., determine the parameters of multipath components (MPC), such as direction of arrival and delay, with resolution better than the Fourier resolution of the setup. We analyze the question both generally, and with respect to a particular novel multi-beam mmWave channel sounder that is capable of performing multiple-input-multiple-output (MIMO) measurements in dynamic environments. 
{\color{black}We firstly propose a novel two-step calibration procedure that provides higher-accuracy calibration data that are required for Rimax or SAGE. Secondly, we investigate the impact of center misalignment and residual phase noise on the performance of the parameter estimator. Finally we experimentally verify the calibration results and demonstrate the capability of our sounder to perform super-resolution parameter estimation.}
\end{abstract}


%
\IEEEpeerreviewmaketitle
\section{Introduction}

Communication in the \gls{mmwave} band will constitute an essential part of 5G communications systems, both for mobile access, as well as fixed wireless access and backhaul \cite{andrews2014will,rappaport2014millimeter}. The design and deployment of such systems requires a thorough understanding of the propagation channel, since important design questions, such as beamforming capability, coverage, and equalizer length, all critically depend on the propagation conditions. 

Understanding of \gls{mmwave} channels is challenging because the physical propagation mechanisms differ significantly from those at lower frequencies. The free-space pathloss increases with frequency (assuming constant antenna gain) \cite{Molisch_book_2000}, and diffraction processes become less efficient. Meanwhile, due to the large bandwidth, and the relatively large antenna arrays used at \gls{mmwave} frequencies, the percentage of delay/angle bins that carry significant energy is low \cite{samimi201328}, which can be interpreted as a sparse structure either in the multipath delay or angular domain. Due to these different propagation conditions, new channel models are required. Several standardization groups, e.g., 3GPP \cite{3gppAbove6GHz} and METIS2020 \cite{metis2020}, have established such models. However, it must be stressed that these models were finalized under time pressure and have the purpose of comparing different systems under comparable circumstances; they are {\em not} suitable for complete understanding of propagation effects or absolute  performance predictions. Several other standardization groups, such as IC1004 \cite{ic1004} and NIST 5G \gls{mmwave} channel model \cite{NIST5G}, are currently developing more detailed models. 

Any improved understandings of \gls{mmwave} channels, as well as the new channel models, rely on thorough and extensive measurements. Since the late 1980s, hundreds of paper have been published describing  \gls{mmwave} channel measurements in different environments, see, e.g., \cite{haneda2015channel} for an overview and references. Since  \gls{mmwave} systems use (massive) MIMO, directionally resolved channel measurements are of special interest.  For outdoor long distance measurements, a popular \gls{mmwave} channel sounding method is based on the use of rotating horn antennas \cite{rappaport2015wideband,hur2014synchronous,kim201528ghz}, i.e., the horn antennas are mechanically pointed into different directions at different times. Depending on the horn beam width, the mechanical rotation of both \gls{tx} and \gls{rx} antennas could potentially take up to several hours. The phase between \gls{tx} and \gls{rx} \glspl{lo} is extremely difficult to preserve over such a long period of time, so that only noncoherent evaluations are possible: data can be evaluated by spectrum peak searching methods, e.g., \cite{bas2017real,akdeniz2014millimeter}. Despite the simplicity and effectiveness in some scenarios, there are obvious drawbacks attributed to the discarding of the phase information, e.g., difficulty to resolve multiple paths whose directions are within the \gls{hpbw} of the horn antenna, and accounting for the sidelobes of imperfect beam patterns, in particular deciding whether any secondary peak is attributed to the sidelobes of the main peak or the contribution from a weak \gls{mpc}. 

In contrast, phase-coherent measurements are possible with (i) setups based on a  \glspl{vna} combined with virtual arrays (mechanical movement of a single antenna) for short-distance measurements, or (ii) real-time \gls{mmwave} multi-beam channel sounders, which may be based on switched horn arrays \cite{papazian2016radio} or phased-arrays with electronically switched beams \cite{bas2017real}. Data can be evaluated by advanced signal processing algorithms that rely on the \gls{mle}. For instance, the \gls{sage} \cite{fleury2002high} is used in Ref. \cite{gustafson2011directional,gustafson2014mm} along with rectangular virtual arrays and in Ref. \cite{Yin_et_al_EuCAP2014} with rotating horns, while Rimax \cite{martinez2014deterministic} was applied in an indoor \gls{mmwave} measurement campaign at \SI{60}{GHz} with \gls{vna} and virtual arrays. {\color{black}Indoor measurements from a switched beam sounder were evaluated with \gls{sage} in Ref. \cite{papazian2016radio}.}


We have recently constructed a real-time \gls{mmwave} multi-beam channel sounder, which is based on phased arrays and electronically switched beams \cite{bas2017real}. {\color{black}The phased array is integrated with other RF elements, such as converters, filters and \glspl{pa}, in one box, which we refer to as RFU for the remainder of the paper. More details about the RFU can be found in Ref. \cite{psychoudakis2016mobile}}. The setup achieves a measurable path loss of \SI{159}{dB} without any averaging or spreading gain, which is sufficient to handle challenging outdoor \gls{nlos} scenarios. More importantly compared to the sounder equipped with rotating horn antennas, our setup is capable to switch between beams in less than $\SI{2}{\mu s}$ with a control interface implemented in \gls{fpga}. As a result, the duration of one \gls{mimo} snapshot is reduced from hours to milliseconds. Along similar lines, {\color{black}Papazian \etal built a \gls{mmwave} sounder that also has a small antenna switching time of $\SI{65.5}{\mu s}$ \cite{papazian2015radio}.} This feature has three main benefits. Firstly it allows the collection of data at tens of thousands of measurement locations in a single measurement campaign; secondly it makes the sounder suitable for measurement campaigns in dynamic environments; thirdly given the small phase drift indicated in \cite[Fig. 9]{bas2017real} the data evaluation with \gls{hrpe} algorithms such as Rimax is feasible.

An essential prerequisite for applying \gls{hrpe} algorithms on \gls{mimo} measurement data is {\em sounder calibration}. Calibration usually consists of two steps: the first is the cable-through calibration, also known as the back-to-back calibration. It mainly serves the purpose to characterize the \gls{tx} and \gls{rx} system frequency response within the operating bandwidth at different gain settings, such that the effects of the channel sounder on the channel impulse response can be compensated. The second step is the antenna array calibration, which is critical for any \gls{hrpe} algorithm that intends to estimate the directions of \glspl{mpc}. By this procedure, it is possible to produce the antenna de-embedded characterization of the propagation channel \cite{Steinbauer_et_al_2001}, which is highly desirable for \gls{mmwave} system simulations where directional horn antennas with different gains and \glspl{hpbw}, or arrays with different shapes and sizes, might be tested \cite{ji2017antenna}. 

A variety of calibration procedures have been used in the literature. The rotating-horn-based sounder from NYU performs back-to-back calibration with two horn antennas pointing towards each other in an anechoic chamber \cite{maccartney2017flexible}; the procedures are relatively simple because the analysis mostly considers the noncoherent response (the \gls{pdp} and the \gls{aps}). The Keysight sounder in \cite{wen2016mmwave} uses power dividers and two separate back-to-back experiments to obtain the system frequency variation, however limited details are available on the calibration of inter-channel phase imbalance or antenna patterns. Papazian \etal present their complete two-step calibration for their \gls{mmwave} sounder in Ref. \cite{papazian2016calibration}, where they first perform the back-to-back calibration after disconnecting the antennas, then use the NIST near-field probe to obtain the phase centers and individual complex antenna patterns. 
Unfortunately these methods cannot be applied to sounders where the antennas are integrated with other RF elements on the same \gls{pcb} - a situation that occurs not only in our sounder, but is of general interest for mm-wave and higher frequencies, as system integration and efficiency considerations often require a joint packaging of these components. 

In this paper, we thus present the calibration procedures and the verification experiments for \gls{mmwave} \gls{mimo} channel sounders based on phased arrays with integrated RF electronics. 
To accommodate the integrated antenna/electronics design and still obtain the system response and frequency-independent array response needed in \gls{hrpe} analysis, we propose a novel two-setup calibration scheme and formulate the extraction and separation of the two responses as an optimization problem.

The main contributions are thus as follows,
\begin{itemize}
 \item we introduce detailed calibration procedures for a beam-switching channel sounder based on a pair of RFUs, which contains the baseline RFU calibration and the multi-gain RFU calibration;
 \item to comply with the data model of Rimax, we formulate the extraction of the frequency-independent array response and the system response as an optimization problem and propose an optimal solution based on the \gls{lra} method;
 \item we analyze the impact of several imperfections of the channel sounder and calibration method on the performance of Rimax evaluations; 
 \item we perform verification experiments in an anechoic chamber with artificially added static reflectors. 
\end{itemize}

This paper is organized as follows. In section \RNum{2} we introduce the signal model of phased arrays and the data model used in the path parameter estimation. 
In section \RNum{3} we introduce calibration procedures that enable the \gls{hrpe} for \gls{mmwave} channel sounders with integrated antenna arrays. Section \RNum{4} discusses the impact of two unavoidable calibration impairments, namely array center misalignment and the residual phase noise, on the performance of Rimax evaluations. Section \RNum{5} presents the results of the verification experiments with a \gls{mmwave} \gls{mimo} sounder in an anechoic chamber. In section \RNum{6} we draw the conclusions. 

%

The symbol notation in this paper follows the rules below:
\begin{itemize}
\item Bold upper case letters, such as $\B{B}$, denote matrices. $\B{B}()$ represent matrix valued functions.
\item Bold lower case letters, such as $\B{b}$, denote column vectors. $\B{b}_j$ is the $j$th column of the matrix $\B{B}$. $\B{b}()$ stands for a vector valued function.
\item Bold symbols $\B{I}$ denote identity matrices. $\B{I}_b$ means that its number of rows equals the length of $\B{b}$.
 \item Calligraphic upper-case letters denote higher dimensional tensors.
 \item $[\B{B}]_{ij}$ denotes the element in the $i$th row and $j$th column of the matrix $\B{B}$.
 \item Superscripts $^\ast$, $^T$ and $^\dagger$ denote complex conjugate, matrix transpose and Hermitian transpose, respectively.
 \item The operators $\vert f(\B{x}) \vert$ and $\lVert \B{b} \rVert$ denote the absolute value of a scalar-valued function $f(\B{x})$, and the L2-norm of a vector $\B{b}$.
 \item The operators $\otimes$, $\odot$ and $\diamond$ denote Kronecker, Schur-Hadamard, and Khatri-Rao products.
 \item The operator $\oslash$ represents the element-wise division between either two vectors, matrices or tensors, and the operator $\circ$ is the outer product of two vectors.
 \item The operators $\lfloor\,\rfloor$ and $\lceil\,\rceil$ are the floor and ceiling functions, respectively. 
\end{itemize}

\section{Signal Model}
\subsection{Phased Arrays}
\label{sect:URA_phase}
\begin{figure}[!t]
  \centering
  \includegraphics[width =\wid\columnwidth]{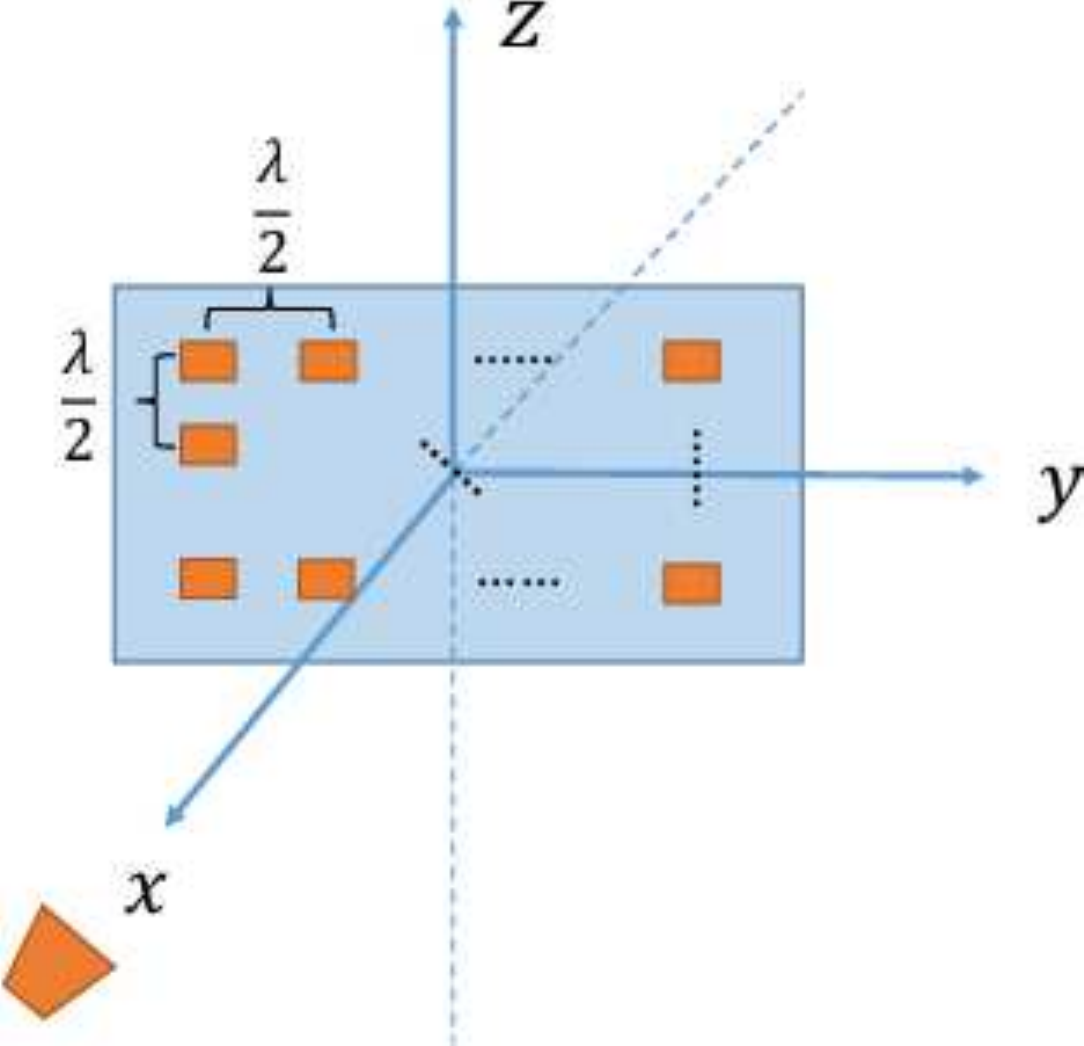}
  \caption{The structure of \gls{ura} with $N_y$ elements along the y-axis and $N_z$ elements along the z-axis}
  \label{fig:URA_2D}
\end{figure}

{\color{black}We consider a \gls{ura} in this paper, which is a popular array configuration for \gls{mmwave} phased arrays}. We assume without loss of generality that the array lies in the y-z plane. Boresight of the array orientation is thus the azimuth angle $\varphi=0^\circ$ and the elevation angle $\theta=90^\circ$, see Fig. \ref{fig:URA_2D}. The array has $N_y$ elements along the y-axis and $N_z$ elements along the z-axis; we denote the total number of antenna elements $N=N_x N_y$. We assume that the antenna elements are separated by $\lambda/2$, i.e. half a wavelength, in both directions. 

We label antennas based on their y-z coordinates, for example we choose the bottom left antenna in Fig. \ref{fig:URA_2D} as the (1,1) element. If the center of the \gls{ura} is aligned with the origin of the cartesian coordinates, the coordinates of the $(n_y,n_z)$ element are 
\begin{align}
  y_{n_y,n_z} &= (n_y - \frac{N_y+1}{2})\frac{\lambda}{2} \label{Eq:antenna_y_coord}\\
  z_{n_y,n_z} &= (n_z - \frac{N_z+1}{2})\frac{\lambda}{2},
\end{align}
where $n_y\in\{x|x\in\mathcal{Z}, 1 \le x \le N_y\}$ and $n_z\in\{x|x\in\mathcal{Z}, 1 \le x \le N_z\}$. We also assume that all antenna elements share the same individual pattern that is given by $A_0(\varphi,\theta)$ with $\varphi\in[-\pi,\pi)$ and $\theta\in[0,\pi]$. Let $\B{X}(\varphi,\theta)$ be a matrix-valued function, i.e., $\B{X}:\mathbb{R}^2\rightarrow \mathbb{C}^{N_y \times N_z}$. It provides the received signal at the \gls{ura} when a plane wave with unit gain arrives from the direction $(\varphi,\theta)$. The function is determined by
\begin{equation}
  [\B{X}(\varphi,\theta)]_{n_y,n_z} = e^{j\B{k}^T\B{p}_{n_y,n_z}}A_0(\varphi,\theta), \label{Eq:URA_th_pattern}
\end{equation}
where $\B{p}_{n_y,n_z}$ is the antenna position vector,
\begin{equation}
  \B{p}_{n_y,n_z}=\begin{bmatrix}
   0 & y_{n_y,n_z} & z_{n_y,n_z}
   \end{bmatrix}^T \label{Eq:antenna_pos_ny_nz}
\end{equation}
and the wave vector \B{k} is given by
\begin{equation}
  \B{k} = \frac{2\pi}{\lambda} \begin{bmatrix}
    \cos\varphi\sin\theta &\sin\varphi\sin\theta &\cos\theta 
    \end{bmatrix}^T.
\end{equation} 

To allow the phased array to steer beams in different directions, we add {\color{black}narrowband} phase shifters to each antenna element. The complex phase weighting matrix \B{W} shares the same dimension with $\B{X}(\varphi,\theta)$, and $[\B{W}]_{n_y,n_z}$ is the corresponding complex weight for the $(n_y,n_z)$-th antenna element. The virtual beam pattern when the \gls{ura} is excited with $\B{W}$ can be mathematically determined by
\begin{equation}
   b_\B{W}(\varphi,\theta) = \tx{tr}\big(\B{W}^T \B{X}(\varphi,\theta)) = \tx{vec}(\B{W})^T \cdot \tx{vec}(\B{X}(\varphi,\theta)), \label{Eq:beam_port}
\end{equation}
where vec() is the vectorization operator. {\color{black}The phase shifters translate the signals from the antenna ports to the beam ports that are associated with different weighting matrices. 
If the phase shifter matrices are DFT (discrete Fourier transform) matrices, the beamports correspond to the beams in the ``virtual channel representation'' of \cite{sayeed2002deconstructing}. We assume that measurements at different beam ports occur at different times, but are all within the coherence time of the channel.} 

\subsection{Rimax Data Model}
\label{sec:Rimax_dataModel}
Similar to the data model in Refs. \cite{richter2005estimation,salmi2009detection}, we use a vector model for the input $T$ \gls{mimo} snapshots, and denote it as $\B{y} \in \mathbb{C}^{M \times 1}$, where $M = M_f\times M_R \times M_T \times T$. It includes contributions from \gls{sp} $\B{s}(\BS\Omega_s)$, \gls{dmc} $\B{n}_\tx{dmc}$ and measurement noise $\B{n}_0$: 
\begin{equation}
  \B{y} = \B{s}(\BS\Omega_s) + \B{n}_\tx{dmc} + \B{n}_{0},  \label{Eq:y_VecModel}
\end{equation}
where the vector $\BS\Omega_s$ represents the parameters of $P$ \gls{sp}s. It consists of polarimetric path weights $\BS\gamma$ and the structural parameters $\BS\mu$ that include $\BS\tau$, $(\BS\varphi_T,\BS\theta_T)$, $(\BS\varphi_R,\BS\theta_R)$, and $\BS\nu$. $\BS\varphi_T$, $\BS\varphi_R$ and $\BS\nu$ are normalized to between $-\pi$ and $\pi$, $\BS\theta_T$ and $\BS\theta_R$ are between $0$ and $\pi$, while $\BS\tau$ is normalized to between $0$ and $2\pi$. The \gls{dmc} follows a Gaussian random process with frequency correlation, and its \gls{pdp} has an exponentially decaying shape \cite{richter2005joint}. The measurement noise is i.i.d. and follows the zero-mean complex Gaussian distribution. {\color{black}The original \gls{sage} \cite{fleury2002high} adopts a similar signal data model except that the \gls{dmc} contribution is neglected.}

An important assumption in the original Rimax algorithm is the array narrowband model, which means the array response is considered \textit{constant} over the frequency band that the channel sounder measures. Although an extension of Rimax to include the wideband array response between \SI{2}{GHz} and \SI{10}{GHz} is included in Ref. \cite{salmi2011propagation}, the narrowband approach has been used almost exclusively in practice; we thus leave wideband calibration for future work. 
Section \ref{sec:calibProc} discusses how to find the best frequency-independent array pattern from the over-the-air calibration data. 

We also assume that the single polarized model is applicable, because the antennas in our phased-array are designed to be vertically polarized. 
The \gls{xpd} of the beam patterns is over \SI{20}{dB} in the main directions. A detailed discussion of the impact of ignoring the additional polarized components in parameter estimation is available in Ref. \cite{landmann2012impact}. 

{\color{black} We assume that a \textit{common} frequency response $\B{g}_f$ attributed to the system hardware is shared between antenna pairs in a \gls{mimo} channel sounder. Meanwhile we adopt the array modeling through \glspl{eadf} \cite{belloni2007doa}, which provides a reliable and elegant approach for signal processing on real-world arrays. The \glspl{eadf} are obtained through performing a \gls{2d} \gls{dft} on the \textit{complex} array pattern either from simulations or array calibration in an anechoic chamber. We denote the \glspl{eadf} for \gls{tx} and \gls{rx} RFUs as $\B{G}_{TV}$ and $\B{G}_{RV}$ respectively. 
Before breaking down the details about $\B{s}(\BS\Omega_{s})$, we first introduce the phase shift matrix $\B{A}(\BS\mu_i) \in \mathbb{C}^{M_i \times P}$ \cite{richter2005estimation}, which is given by
\begin{equation}
 \B{A}(\BS\mu_i) = \begin{bmatrix}
   e^{j(-\frac{M_i-1}{2})\mu_{i,1}} &\cdots & e^{j(-\frac{M_i-1}{2})\mu_{i,P}} \\
    \vdots                                  &           & \vdots \\
   e^{j(\frac{M_i-1}{2})\mu_{i,1}}  &\cdots & e^{j(\frac{M_i-1}{2})\mu_{i,P}}
 \end{bmatrix}.
\end{equation}
$\BS\mu_i$ is a structural parameter vector that represents either $\BS\tau$, $\BS\varphi_T$, $\BS\theta_T$, $\BS\varphi_R$, $\BS\theta_R$ or $\BS\nu$. 
These quantities will be essential for HRPE evaluations.}

Based on the system response and the \glspl{eadf} introduced above, we obtain the basis matrices:
\begin{align}
 \B{B}_f &= \B{G}_f \cdot \B{A}(-\BS\tau)\\
 \tilde{\B{B}}_{TV} &= \Big[ \B{G}_{TV} \cdot \big( \B{A}(\BS\theta_T) \diamond \B{A}(\BS\varphi_T)\big) \Big] \\
 \tilde{\B{B}}_{RV} &=  \Big[ \B{G}_{RV} \cdot \big( \B{A}(\BS\theta_R) \diamond \B{A}(\BS\varphi_R)\big)\Big] \label{Eq:B_RV}\\
 \B{B}_t &= \B{A}(\BS\nu). \label{Eq:B_t}
\end{align}
$\B{G}_f$ is a diagonal matrix with its diagonal elements given by $\B{g}_f$.
With phased arrays we use the beam ports instead of the antenna ports, hence $M_T$ and $M_R$ become the number of beam ports and reflect how many different beamforming matrices $\B{W}$ are applied to \gls{tx} and \gls{rx} arrays respectively. Finally the signal data model for the responses of \gls{sp}s is given by
\begin{equation}
  \B{s}(\BS{\Omega}_{s}) = \B{B}_t \diamond \B{B}_{TV} \diamond \B{B}_{RV} \diamond \B{B}_f \cdot \BS{\gamma}_{VV} 
\end{equation}
{\color{black}If the measurement environment is static such as an anechoic chamber, and one measurement snapshot $\B{y}$ contains only one \gls{mimo} snapshot, i.e. $M_T \times M_R$ pairs of sweeping-beam measurements, the signal model of \glspl{sp} can be simplified to}
\begin{equation}
  \B{s}(\BS{\Omega}_{s}) = \B{B}_{TV} \diamond \B{B}_{RV} \diamond \B{B}_f \cdot \BS{\gamma}_{VV}
\end{equation}

\section{Calibration Procedures}
\label{sec:calibProc}
\begin{figure}[!t]
  \centering
    \centering 
    \subfigure[Over-the-air verification/measurements]{\includegraphics[width = \wid\columnwidth, 
    clip = true]{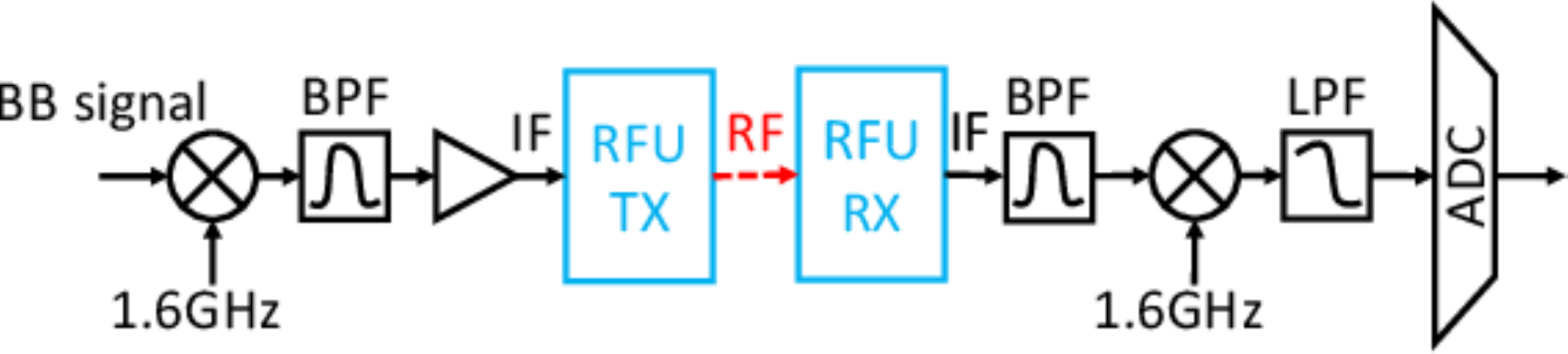}\label{subfig:ota_diagram}} \\
    \subfigure[Through-cable back-to-back calibration]{\includegraphics[width = \wid\columnwidth, 
    clip = true]{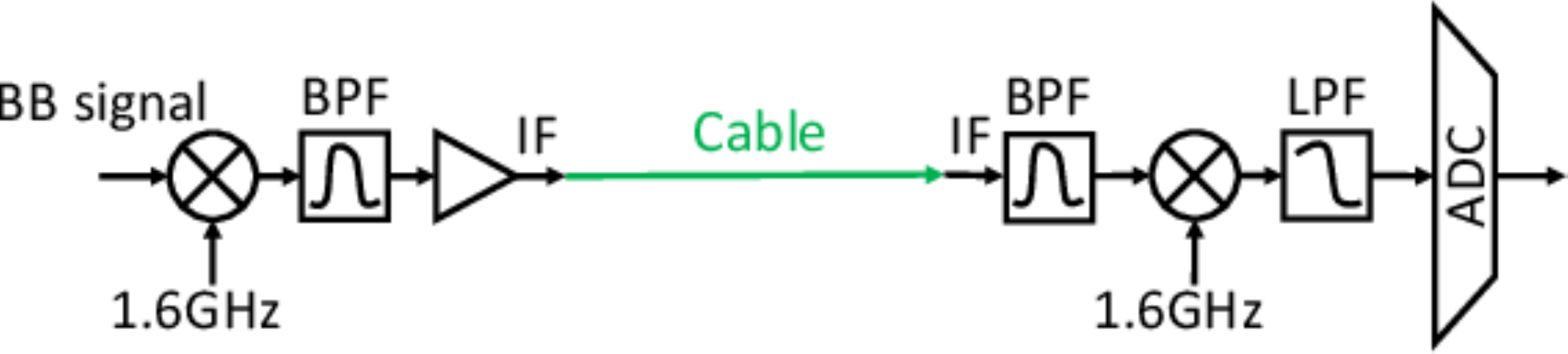}\label{subfig:thru_diagram}} \\
    \subfigure[Over-the-air RFU calibration]{\includegraphics[width = 0.4\columnwidth, 
    clip = true]{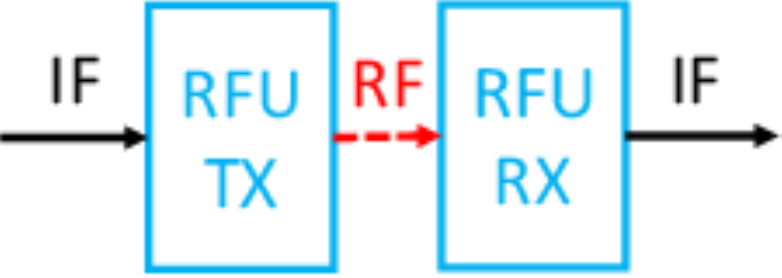}\label{subfig:ota_RFU_diagram}}
    \caption{The system diagrams for the RFU calibration and verification experiments}
  \label{fig:summary_diagram}
\end{figure}

In this section, we describe a novel calibration scheme for mmWave channel sounders with phased-arrays. 
The simplified diagram of the time-domain setup is given in Fig. \ref{fig:summary_diagram}(a), which is the measurement setup for the phase stability test in Section \ref{sec:calibLimit} and verification measurements in Section \ref{sec:Meas}. The overall calibration procedures can also be categorized into a back-to-back calibration in Fig. \ref{fig:summary_diagram}(b) and the RFU/antenna calibration in Fig. \ref{fig:summary_diagram}(c). 

{\color{black}Conventional array calibration is based on the assumption that the antenna or beam ports can be connected to RF signals at the operating frequency, and typically the array to be calibrated would consist only of passive components. In contrast, we consider here the situation that the RFU has an amplifier as well as an embedded mixer and an \gls{lo} at \SI{26}{GHz}, so that the input and output frequencies are different. The \gls{if} frequency is between \SI{1.65}{GHz} and \SI{2.05}{GHz}. This motivates us to use \textit{both} \gls{tx} and \gls{rx} RFUs in the antenna calibration, so that the generator and receiver of the calibration signal in a \gls{vna} can operate on the same frequency (even though it is lower than the operating frequency of the array). Although the measured frequency response based on the setup in Fig. \ref{subfig:ota_RFU_diagram} varies significantly over the \SI{400}{MHz} frequency band, we still attempt to find the best narrowband fit of the RFU pattern to the calibration data, in order to match the assumption of the data model introduced in Section \ref{sec:Rimax_dataModel}.}

In this subsection we introduce the main procedures of RFU calibration. Here we limit our objective to calibrating the frequency and beam pattern responses of one \gls{tx} and one \gls{rx} RFU.\footnote{{\color{black}The calibration procedures can be applied to the setup where both the \gls{tx} and \gls{rx} have multiple panels.}} The procedures consist of two main steps
\begin{enumerate}
  \item the baseline calibration of the \gls{tx} and \gls{rx} RFUs, i.e., calibration of the gain pattern and frequency response for one setting of the amplifier gains at the \gls{tx} and \gls{rx};
  \item the PNA-assisted multi-gain calibration of the \gls{tx} and \gls{rx} RFUs.
\end{enumerate}

The main motivation for a two-step calibration is that - for reasons explained below -  the {\em complex} phase pattern can only be extracted for the combination of TX and RX. On the other hand, performing such a joint calibration for all possible {\em combinations} of gain settings at TX and RX is practically infeasible due to the excessive time it would take, so that the multi-gain calibration has to be done separately for \gls{tx} and \gls{rx}, with the help of a different setup.

\begin{figure}[!t]
   \centering
   \subfigure[The baseline calibration]{\frame{\includegraphics[width = 0.4\columnwidth]{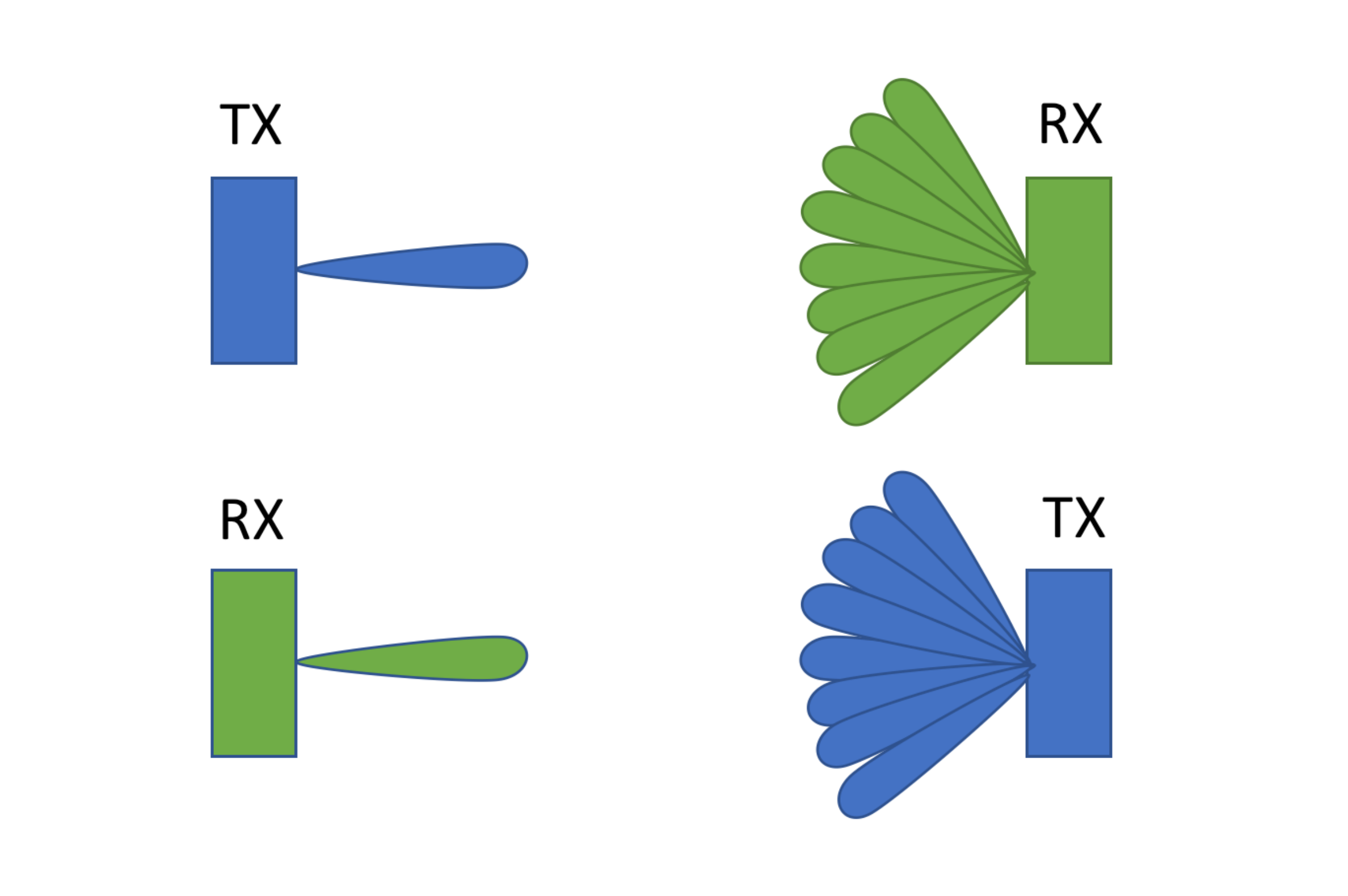}}
   \label{fig:Setup_baseline}}
   \subfigure[The PNA-assisted multigain calibration]{\frame{\includegraphics[width = 0.4\columnwidth]{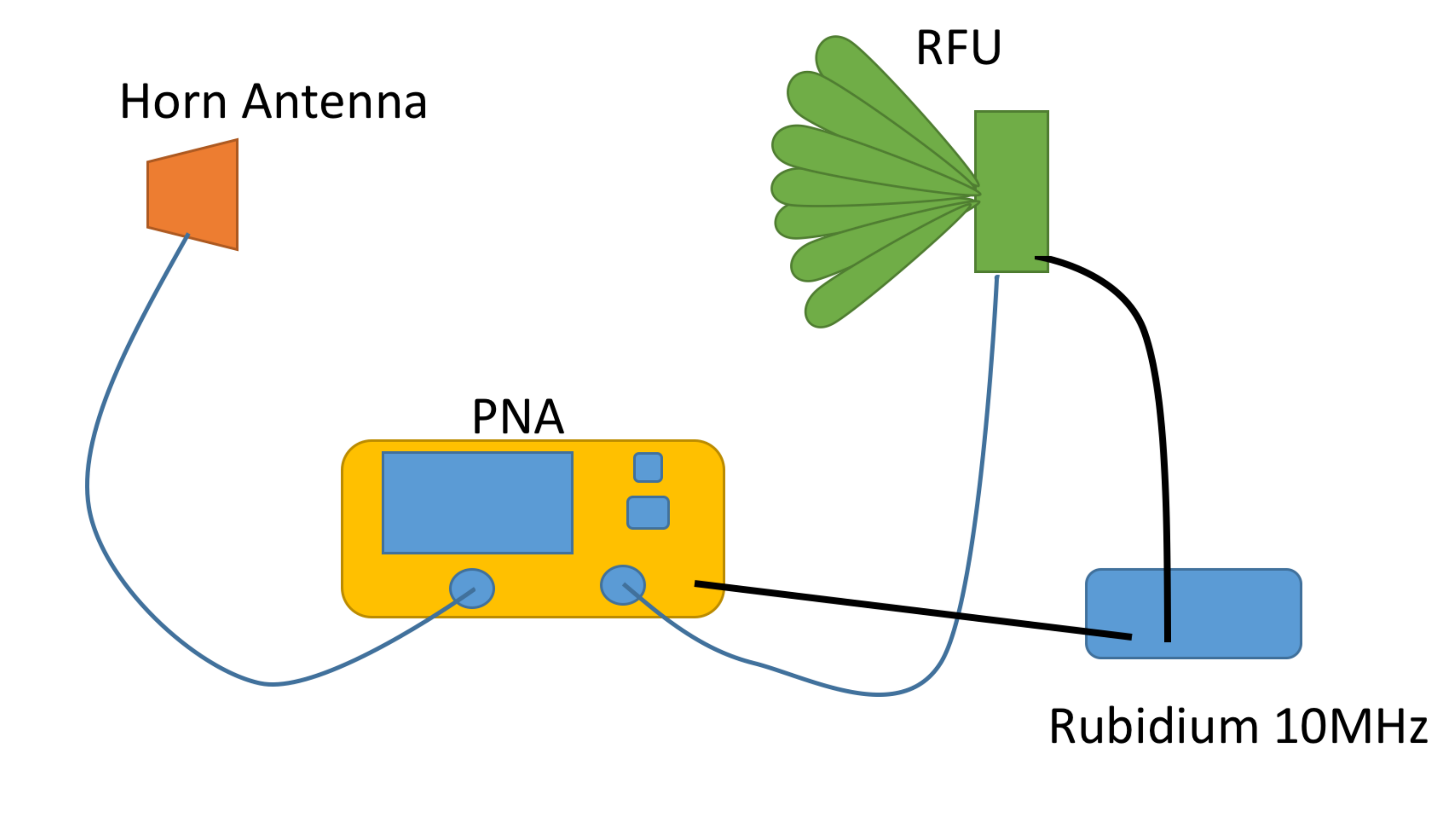}}
   \label{fig:Setup_multigain}}
   \caption{The setup diagrams of the RFU calibration in an anechoic chamber with birdview} 
\end{figure}

To simplify the discussion, we only consider calibration as a function of azimuth; elevation can treated similarly. The measurement setup is illustrated in Fig. \ref{fig:Setup_baseline}. The same Rubidium reference is shared between the two RFUs and the VNA (in our experiments, the VNA we use is Keysight model KT-8720ES). The main procedures are given as follows. In the first experiment we treat the \gls{tx} RFU as the probe and the \gls{rx} RFU as the \gls{dut}, so we fix both the orientation and the beam setting of the \gls{tx} RFU while placing the \gls{rx} RFU on a mechanical rotation stage and turning it to different azimuthal orientations and measuring with the \gls{vna}. The \gls{vna} outputs S21 responses at different \gls{rx} phase shifter settings (i.e., beams) and different azimuth angles, which are denoted as $Y_1(n,\varphi_R,f)$, where the beam index $n=1,2,\ldots,19$. The calibration sequence for these three dimensions basically follows embedded FOR loops, and the loop indices from inside to outside are $f \rightarrow n \rightarrow \varphi_R$. This sequence is motivated by the fact that scanning through frequencies is faster than scanning through beams (which requires phase shifter switching, which takes a few microseconds), which in turn is faster than scanning through observation angles, which requires mechanical rotation and thus a few seconds. Shortening the measurement time is not only a matter of convenience, but also reduces the sensitivity to inevitable phase noise, see Section \ref{sec:calibLimit}.                                                                                                                                                                                                                                                                                                                                                                                                       Similarly after swapping the positions and the roles of the \gls{tx} and \gls{rx} RFUs, we perform the second experiment that generates $Y_2(m,\varphi_T,f)$ with the \gls{tx} beam index $m=1,2,\ldots,19$. 

The goal of the baseline calibration is to estimate the frequency-independent beam patterns $B_T(m,\varphi_T)$ and $B_R(n,\varphi_R)$ from $Y_1$ and $Y_2$. We can build a joint estimator by minimizing the sum of squared errors, and the corresponding optimization problem is stated as
\begin{align}
 \underset{B_T,B_R,G_0,k}{\textbf{min}}\; &\sum \,\lvert B_{T0}B_R(n,\varphi_R)G_0(f)-Y_1(n,\varphi_R,f) \rvert^2 \notag \\ &+ \lvert kB_{R0}B_T(m,\varphi_T)G_0(f) -Y_2(m,\varphi_T,f) \rvert^2, 
 \label{Prob:Jt_EADFExt}
\end{align}
where $k$ is a complex scalar that attempts to model the gain and reference phase offset between two experiments illustrated in Fig. \ref{fig:Setup_baseline}. The gain difference is due to the small boresight misalignment and the phase offset is because of different phase values in \glspl{lo} at the start of the two experiments. Besides the gains of two probes are given by $B_{T0}\triangleq B_T(10,0)$ and $B_{R0} \triangleq B_R(10,0)$.

To simplify the problem formulation, we use the vector notation. For example $\B{b}_R \triangleq \tx{vec}(\textbf{\text{B}}_R)$, $\B{b}_T \triangleq \tx{vec}(\B{B}_T)$ and $\B{g}_f$ is the frequency vector related to $G_0(f)$. We also need to transform the two data sets into matrices, $\B{Y}_1 \triangleq  \tx{reshape}(Y_1(n,\varphi_R,f),[\,],N_f)$ and $\B{Y}_2 \triangleq  \tx{reshape}(Y_2(m,\varphi_T,f),[\,],N_f)$. Here reshape() is the standard MATLAB function. The original problem in (\ref{Prob:Jt_EADFExt}) can then be rewritten as 
\begin{align}
   \underset{\B{b}_R,\B{b}_T,\B{g}_f,k}{\textbf{min}}\; \lVert B_{T0} \B{b}_R \B{g}_f^\tx{T}  - \B{Y}_1 \rVert_F^2 + \lVert kB_{R0} \B{b}_T \B{g}_f^\tx{T} - \B{Y}_2 \rVert_F^2. \label{Prob:Jt_EADFExt_org}
\end{align}
If we combine the two data sets into one, we can have $\B{Y}=[\B{Y}_1;\B{Y}_2]$. The problem in (\ref{Prob:Jt_EADFExt_org}) is equivalent to
\begin{align}
   \underset{\B{u},\B{v}}{\textbf{min}}\; \lVert \B{u}\B{v}^\dagger - \B{Y} \rVert_F^2, \label{Prob:Jt_EADFExt_vec}
\end{align}
which is a typical \gls{lra} problem and it can be efficiently solved through \gls{svd} of \B{Y} \cite{eckart1936approximation}. If we denote its optimal solutions as $\B{u}^\circ$ and $\B{v}^\circ$, we easily find the mapping from $\B{u}^\circ$ to $\B{b}_T$, $\B{b}_R$ and k, as well as the mapping from $\B{v}^\circ$ to $\B{g}_f$. The optimal solution to (\ref{Prob:Jt_EADFExt_vec}) can be found through Alg. \ref{Alg:Jt_EADFExt_vec}.
\begin{algorithm}[!ht]
 \caption{The SVD-based algorithm to solve the problem (\ref{Prob:Jt_EADFExt_vec})}
 \label{Alg:Jt_EADFExt_vec}
 \begin{algorithmic}[1]
   \STATE Stack $\B{Y}_1$ and $\B{Y}_2$ in the rows, $\B{Y}=[\B{Y}_1;\B{Y}_2]$
   \STATE Perform the \gls{svd} on $\B{Y}=\B{U}\BS\Sigma\BS{V}^\dagger$ and find the largest singular value $\sigma_1$ and its related singular vectors $\B{u}_1$ and $\B{v}_1$ 
   \STATE Initialize $\B{g}_f = \sigma_1\B{v}_1^\ast$;  Divide $\B{u}_1$ into two halves with equal length  $\B{u}_1=[\B{u}_{1,1};\B{u}_{1,2}]$
   \STATE Finally $\B{b}_R=a\B{u}_{1,1}$; $a$ is selected such that the center element of $\B{b}_R$ is 1
   \STATE $\BS{\tilde{u}}_{1,2} = a \B{u}_{1,2}$, the final $\B{g}_f=\frac{1}{a}\B{g}_f$ 
   \STATE Finally $k$ equals the center element of $\BS{\tilde{u}}_{1,2}$,  the final $\B{b}_R=\BS{\tilde{u}}_{1,2}/k$
 \end{algorithmic}
\end{algorithm}
We also normalize the probe gains by setting $B_{T0}$ and $B_{R0}$ to 1. 

\begin{figure}[!t]
 \centering
 \includegraphics[width = \wid\columnwidth]{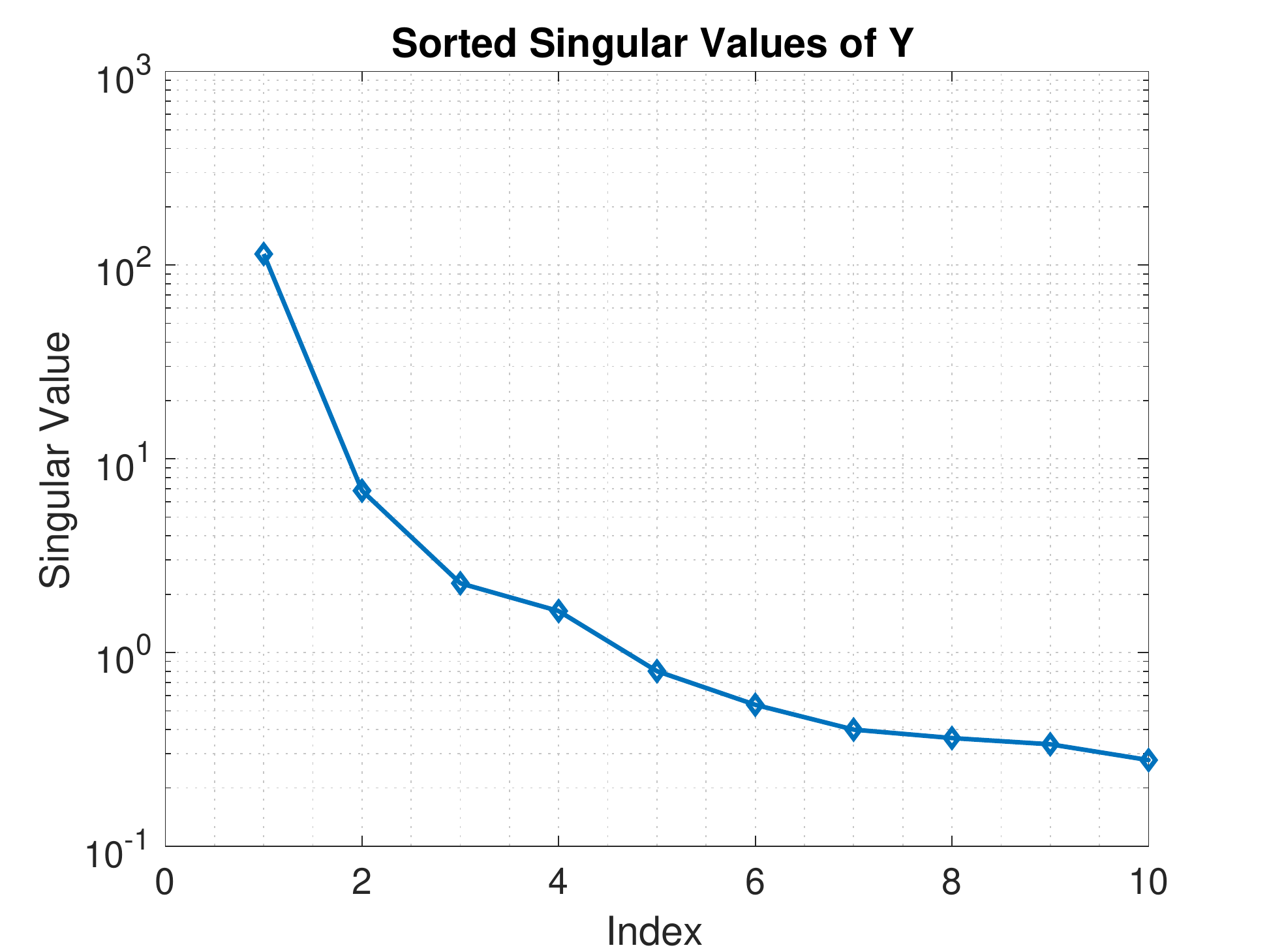}
 \caption{{\color{black}The distribution of the ten largest singular values of $\B{Y}=[\B{Y}_1;\B{Y}_2]$ in Alg. \ref{Alg:Jt_EADFExt_vec}.}}
 \label{fig:sort_sValue}
\end{figure}

\begin{figure}[!t]
  \centering
  \subfigure[For $Y_1$ when the \gls{if} $f=\SI{1.85}{GHz}$]{\includegraphics[width = 0.45\columnwidth]{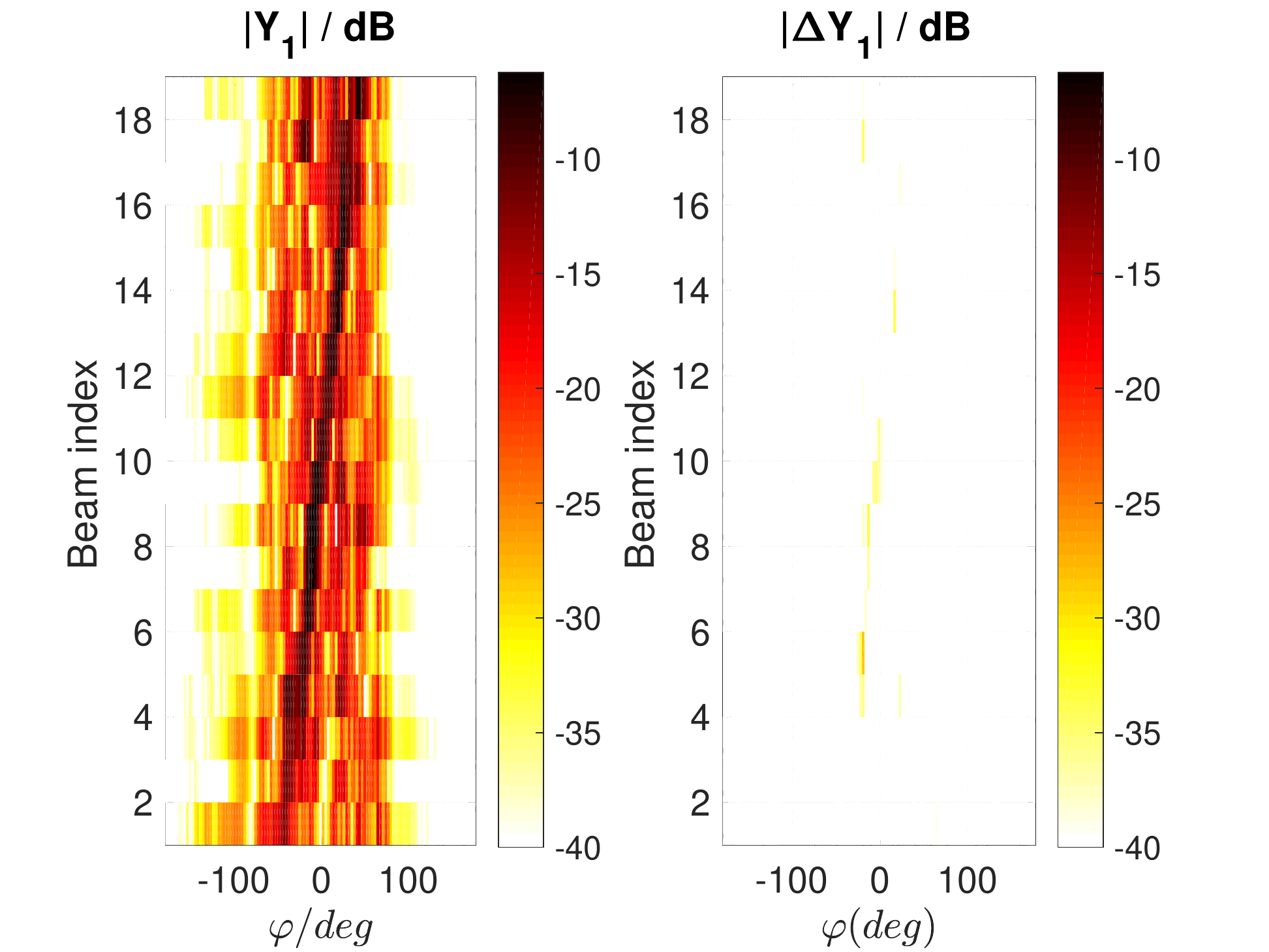}
  \label{fig:Y_1_comp_f}}
  \subfigure[For $Y_2$ when the beam index $n=8$]{\includegraphics[width = 0.45\columnwidth]{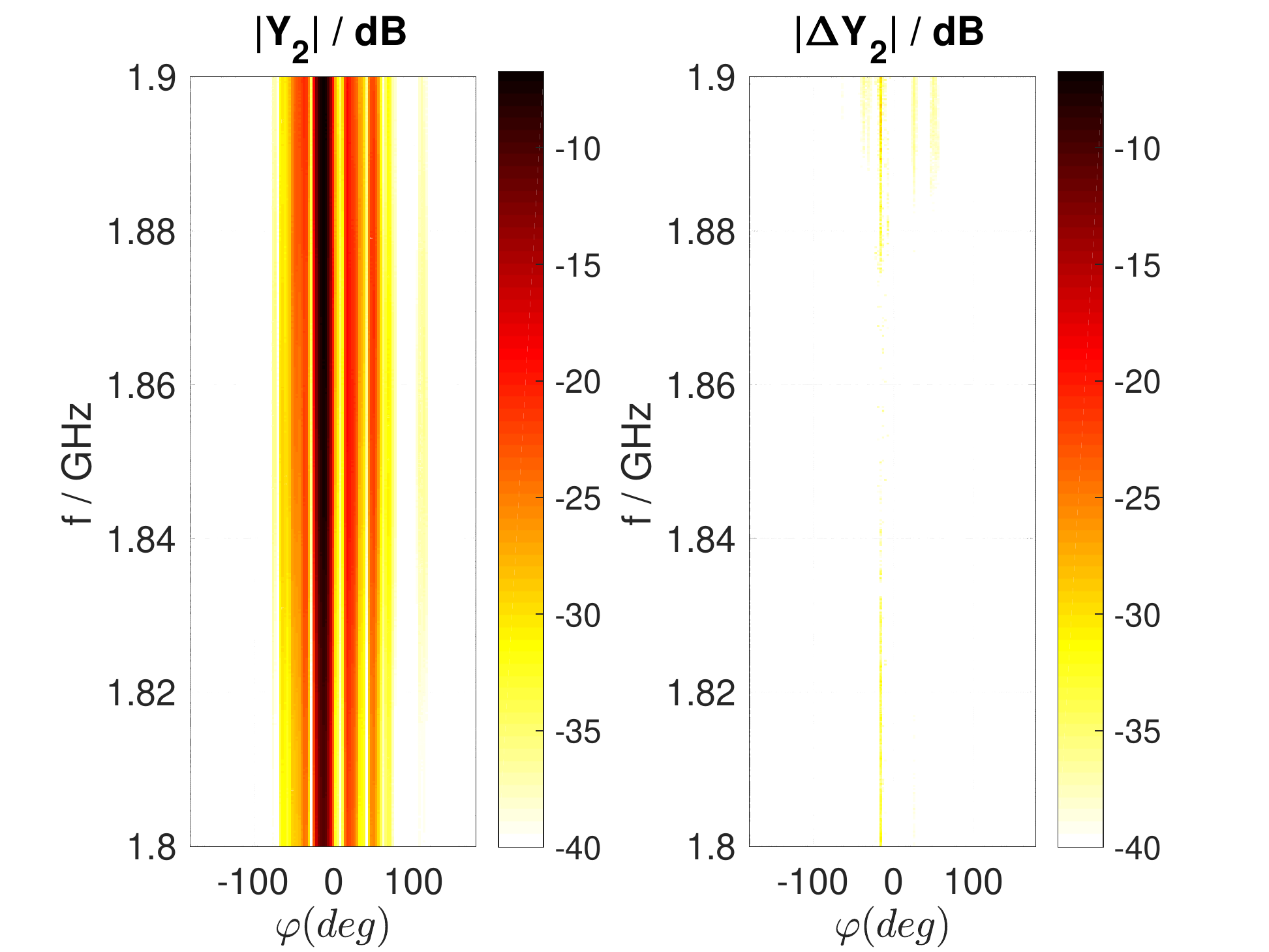}
  \label{fig:Y_2_comp_b}}
  \caption{The comparison between $Y$ and reconstructed $\hat{Y}$ to demonstrate the goodness-of-fit for the pattern extraction algorithm in Alg. \ref{Alg:Jt_EADFExt_vec}.}
\end{figure}

After implementing and testing the algorithm on actual calibration data with \SI{100}{MHz} bandwidth, we have obtained quite good pattern extraction results\footnote{The larger the processing bandwidth is, the worse the agreement becomes between the assumed model and the data, which negatively impacts the accuracy of the \gls{hrpe}. On the other hand a larger bandwidth could help improving the resolution of delay estimation. The determination of the optimal bandwidth is out of the scope of this paper.}. The sorted singular values of $\B{Y}$ are shown in Fig. \ref{fig:sort_sValue}, where the ratio between the largest and the second largest singular value is 16.7. {\color{black}For the ideal solution there should only be one nonzero eigenvalue.} With the solutions to (\ref{Prob:Jt_EADFExt_vec}), we compare $Y_1(n,\varphi,f)$ against $\hat{Y}_1$ in Fig. \ref{fig:Y_1_comp_f} while fixing $f=\SI{1.85}{GHz}$. Similarly the comparison results of $Y_2$ can be found in Figs. \ref{fig:Y_2_comp_b} when we fix $n=8$. The complex scalar $k$ is $0.078-1.045i$, and its amplitude is about \SI{0.40}{dB}.

We now turn to the second step of the calibration. In order to calibrate the RFU responses at different gain settings, we need to have a calibration setup that can handle the gain variations at both \gls{tx} and \gls{rx} RFUs without saturating any device. The requirement is quite difficult to fulfill in the baseline calibration setup highlighted in Fig. \ref{fig:Setup_baseline}, therefore we propose to perform a second multi-gain calibration procedure via using a PNA for \gls{tx} and \gls{rx} separately. The setup is illustrated in Fig. \ref{fig:Setup_multigain}. It uses the measurement class known as the frequency converter application (Option 083) in the Keysight PNA series. We use the configuration known as ``SMC + phase" which provides good measurement results on the conversion loss and the group delay of the \gls{dut}. {\color{black}To simplify the task of measuring the conversion loss and the phase response of a mixer, Dunsmore \cite{dunsmore2011new} introduces a new calibration method on a VNA by using a phase reference, such as a comb generator traceable to NIST, to calibrate the input and output phase response of a VNA independently, which eliminates the need for either reference or calibration mixer in the test system.
} However we \emph{cannot} measure with the \gls{vmc} configuration, which could have provided the reference phase apart from the group delay, because finding reciprocal calibration mixers that match the \gls{if} and the mmWave RF frequency is very difficult. The quality of the calibration mixers ultimately limits the performance of the \gls{vmc} method \cite{KT-FCAaccuracy}. {\color{black}Although we do not have a direct control of the \gls{lo} into the RFU, we can share the \SI{10}{MHz} reference clock of the PNA with the RFU, so that the phase response from the PNA is also stable.} 

The objectives of this multi-gain calibration are twofold: (i) validating the frequency-independent patterns in the baseline calibration\footnote{We expect the frequency-independent \glspl{eadf} implicitly required by Rimax to hold well at different RFU gain settings.}; (ii) calibrating and estimating the frequency responses of the \gls{tx} and \gls{rx} RFUs separately. As shown in \cite[Fig. 8]{bas2018real_arxiv}, the variable gain controller affects the signal power at \gls{if} for both \gls{tx} and \gls{rx} RFUs.
The calibration procedures for the setup in Fig. \ref{fig:Setup_multigain} are given as follows. We align a standard gain horn antenna with the boresight of the \gls{rx} RFU, then measure the S21 for the \gls{rx} RFU with different beam configurations, different gain settings before we rotate the RFU to the next azimuth angle. The same steps are repeated for the \gls{tx} RFU, except that we reverse the input and output signals when the horn antenna becomes the receiving antenna. These two steps produce two data sets, which we denote as $Y_R(g,n,\varphi,f)$ and $Y_T(g,n,\varphi,f)$, where the 4-tuple $(g,n,\varphi,f)$ represents the RFU \gls{if} gain setting, the beam index, the azimuth angle and frequency index. Similarly to the baseline calibration setup, the calibration procedures here closely follow embedded FOR loops. and the loop indices from inside to the outside are $f \rightarrow n \rightarrow g \rightarrow \varphi$. An important feature about the two data sets is that they are only phase coherent within the same frequency sweep, because of the random initial phase of each sweep in the ``SMC+phase'' calibration configuration. For this reason, they cannot be used for the baseline calibration.

First we remove the responses of the standard gain horn and the \gls{los} channel from $Y_R$ or $Y_T$.
To estimate different frequency responses and verify the magnitude of RFU patterns extracted in the baseline calibration, we formulate the following optimization problem,
\begin{align}
 \underset{\B{b},\B{g}}{\textbf{min}}\; &\lVert \B{b}\B{g}^\tx{T} - \B{Y}_{R} \rVert_F^2 \label{Prob:MultiGain_est} \\
 \tx{s.t.}\; & |\B{b}| = \B{b}_a. \notag
\end{align}
This problem formulation tries to find the best rank-1 approximation to $\B{Y}_{R}$ subject to the constraint that the \textit{magnitude} of the optimal \B{b} is equal to the \textit{amplitude} pattern $\B{b}_a$ extracted from the baseline calibration. {\color{black}Comparing to the problem in (\ref{Prob:Jt_EADFExt_vec}), the vector equality constraint prevents us from applying Alg. \ref{Alg:Jt_EADFExt_vec}.} However we can substitute \B{b} with $\B{b}_a \odot e^{j\BS\phi}$ in the objective function, and propose an iterative algorithm based on the alternating projection method in Alg. \ref{Alg:MultiGain_EstVerify}. The algorithm attempts to solve two smaller sub-problems iteratively until the solution converges. The two matrices in the subproblems are given by $\B{A}_b = \B{I}_g \otimes (\B{b}_a \odot e^{j\BS\phi})$ and $\B{A}_g = \B{g} \otimes \B{I}_b$.

\begin{algorithm}[!ht]
 \caption{The iterative optimization algorithm to solve the problem (\ref{Prob:MultiGain_est})}
 \label{Alg:MultiGain_EstVerify}
  \begin{algorithmic}[1]
    \STATE Find initial estimates for the pattern phase vector $\BS\phi$ and frequency response $\B{g}$;
    \WHILE {$\|(\B{b}_a \odot e^{j\BS\phi})\B{g}^\tx{T} - \B{Y} \|_F^2$ has yet converged}
    \STATE Fix $\BS\phi$, and use the least-square method to solve the sub-problem 1: \\
    $\min_{\B{g}} \|\mathbf{\B{A}_b\B{g}-\text{vec}\{\mathbf{Y}\}}\|^2$;
    \STATE Fix $\mathbf{g}$, and use the Levenberg-Marquardt method \cite{marquardt1963algorithm} to solve the nonlinear optimization sub-problem 2: \\
    $\min_{\BS\phi} \|\B{A}_g(\B{b}_a\odot e^{j\BS\phi})-\text{vec}\{\mathbf{Y}\}\|^2$.
    \ENDWHILE
  \end{algorithmic}
\end{algorithm}

Firstly we process $Y_{R/T}$ when the RFU gain setting equals that in the baseline calibration, so that we can obtain the frequency response from \gls{tx} and \gls{rx} RFUs separately, because only one RFU is involved in the PNA-aided gain calibration. However the product of these two frequency responses should in principle be close to $\B{g}_f$ extracted from Alg. \ref{Alg:Jt_EADFExt_vec}, which serves as a part of the verifications.  
Secondly we repeat the steps in Alg. \ref{Alg:MultiGain_EstVerify} for different \gls{tx} and \gls{rx} RFU gain settings. {\color{black}Among them we select the set of gain settings whose mismatch errors, evaluated according to the objective function in (\ref{Prob:MultiGain_est}), are relatively small, and we consider using these gain settings in future measurements.} 

\section{{\color{black}Calibration Practical Limitations}}
\label{sec:calibLimit}
This section investigates two important practical issues in the mmWave calibration procedure. The first is the misalignment between the calibration axis and the center of the antenna array. The second is the phase stability measurement of two RFUs in the anechoic chamber, and we observe that the residual phase noise is composed of a slow-varying component and another fast-varying term. 
We study the impact of these issues on the performance of Rimax evaluation with simulations. 

\subsection{Center Misalignment}
It is important to align the rotation axis with the phase center of an antenna in the antenna calibration in the anechoic chamber. Different methods have been proposed to calculate the alignment offset based on the phase response of the calibration data \cite{prata2002misaligned,chen2014determining}. 

We again consider the \gls{ura} shown in Fig. \ref{fig:URA_2D}. Let us assume that the origin of the Cartesian coordinates is  aligned with the center of the \gls{ura}. 
The probe horn antenna is placed at $\B{p}_{t}=[5,0,0]^T$. The antenna position vector with the \textit{ideal} alignment is given by (\ref{Eq:antenna_pos_ny_nz}) for the $n_y$-th and the $n_z$-th element. We assume that this is the initial position with $\varphi=0^\circ$ and $\theta=90^\circ$ of the array pattern calibration. If we denote the offset vector at the initial position as $\Delta\B{p}$, the actual initial position is given by $\tilde{\B{p}}_{n_y,n_z}=\B{p}_{n_y,n_z}+\Delta\B{p}$. 

To measure the array at $\varphi_0$ and $\theta_0$ we can compute the new antenna position with the rotation matrix, which is given by
\begin{equation}
  \tilde{\B{p}}_{n_y,n_z}(\varphi_0,\theta_0) = \B{R}_y(\theta_0-90)\B{R}_z(\varphi_0)\tilde{\B{p}}_{n_y,n_z}. 
\end{equation}
where $\B{R}_y$ and $\B{R}_z$ are the standard $3 \times 3$ transformation matrices that represent rotation along the y axis with the right-hand rule and the z axis with the left-hand rule respectively \cite{strang1993introduction}. The distance between the probe and the rotated antenna is given by $d_{n_y,n_z}(\varphi_0,\theta_0)=\lVert \tilde{\B{p}}_{n_y,n_z}(\varphi_0,\theta_0)-\B{p}_t \rVert_2$. Therefore the simulated ``distorted'' calibration response is given by
\begin{equation}
  \tilde{b}_{n_y,n_z}(\varphi_0,\theta_0) = A_{n_y,n_z}(\varphi_0,\theta_0)e^{-j2\pi f_c \frac{d_{n_y,n_z}}{c_0}}, 
\end{equation}
where $f_c$ is the carrier frequency, $c_0$ is the speed of light in air, and $A_{n_y,n_z}(\varphi_0,\theta_0)$ is the element pattern. For simplicity we assume in the simulations that antennas are isotropic radiators, $\ie\,A_{n_y,n_z}(\varphi_0,\theta_0)=1$. 
Similarly we could acquire the \textit{ideal} pattern $b_{n_y,n_z}(\varphi_0,\theta_0)$ via setting $\Delta\B{p}$ as $\B{0}$. 
Examples are shown in Fig. \ref{fig:EADF_comp_misalign}. As the offset $\Delta\B{p}$ increases we can observe that the high power coefficients in \glspl{eadf} tend to be more spread-out when compared to the ideal case in Fig. \ref{subfig:EADF_ideal}. This will decrease the effectiveness of mode gating, where we could truncate the coefficients in \gls{eadf} in order to reduce the calibration noise. {\color{black}Landmann \etal suggest estimating the phase drift due to the center misalignment together with \glspl{eadf} from the array calibration data \cite{landmann2006estimation}. However this method becomes less effective in our case when the calibration is further affected by the fast phase variation, see Section \ref{sect:PN_model}.}

\begin{figure}[!t]
  \centering
  \includegraphics[width = \wid\columnwidth]{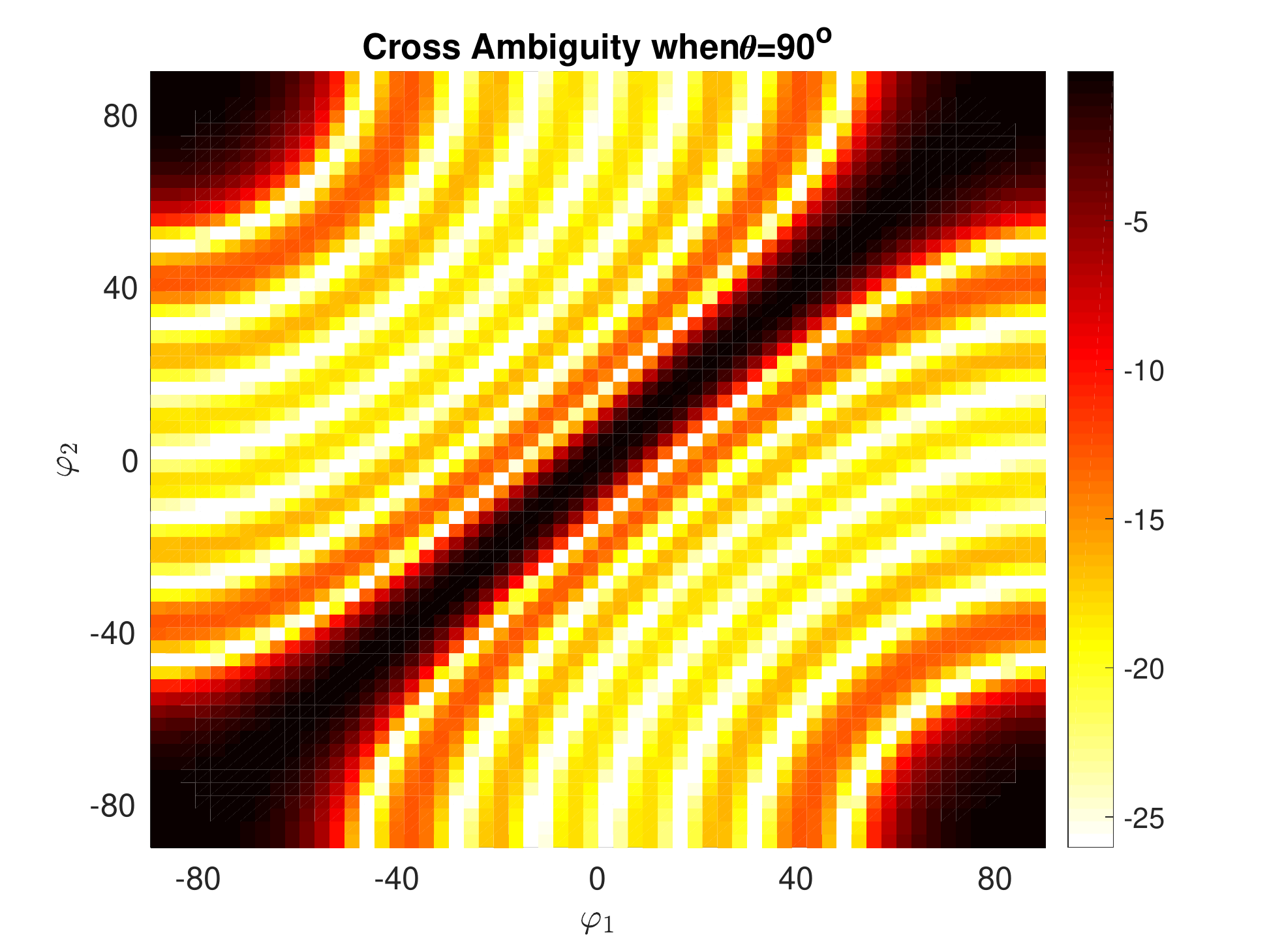}
  \caption{The cross ambiguity function between $\B{b}$ and $\tilde{\B{b}}$ when $\theta_1=\theta_2=90^\circ$}
  \label{fig:XAmbFunc}
\end{figure}

\begin{figure}[!t]
  \centering
  \subfigure[$\Delta\B{p}=\B{0}$]{\includegraphics[width = 0.4\columnwidth, 
  clip = true]{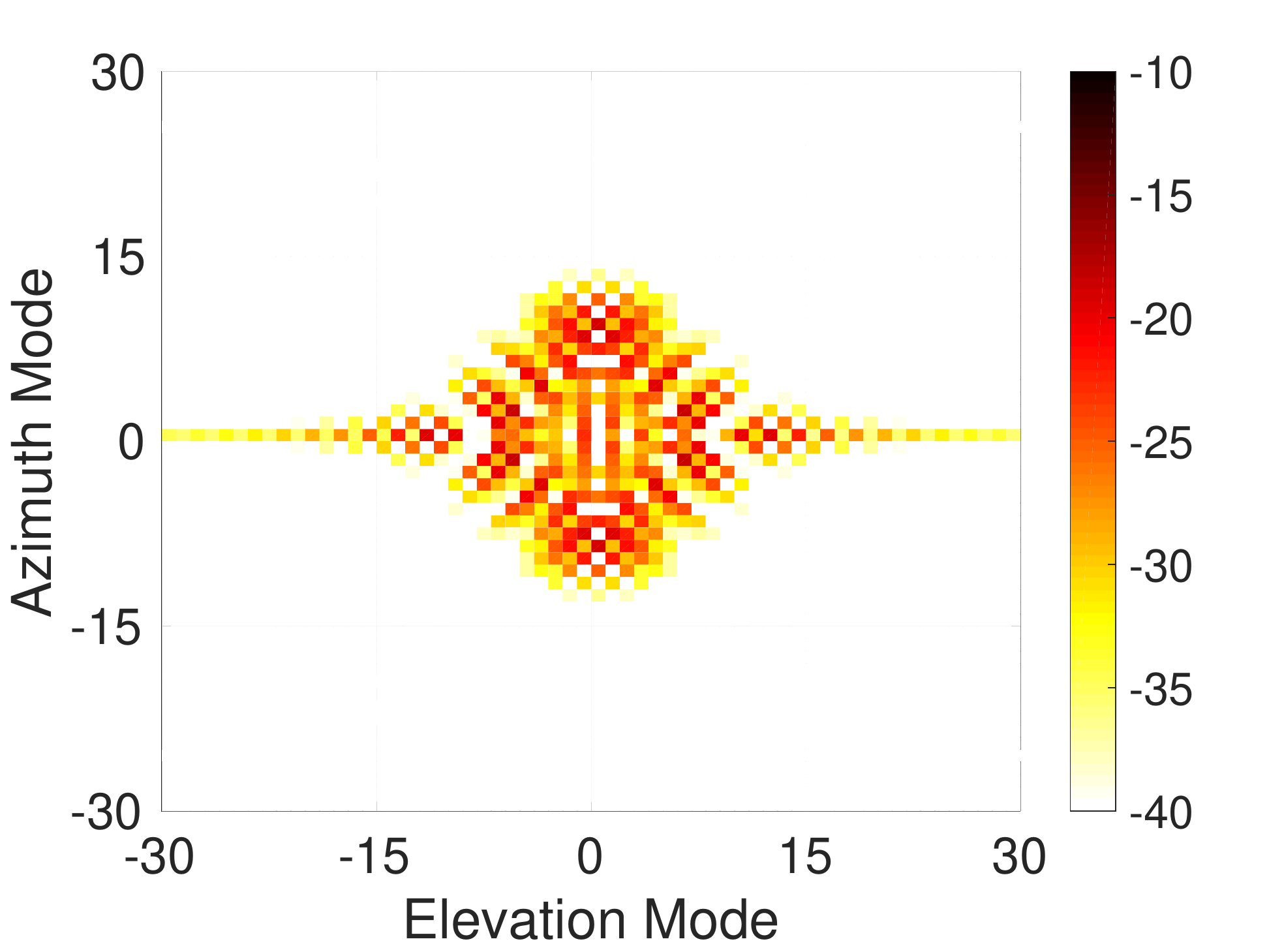}\label{subfig:EADF_ideal}}
  \subfigure[$\Delta\B{p}=\lambda\B{1}$]{\includegraphics[width = 0.4\columnwidth, 
  clip = true]{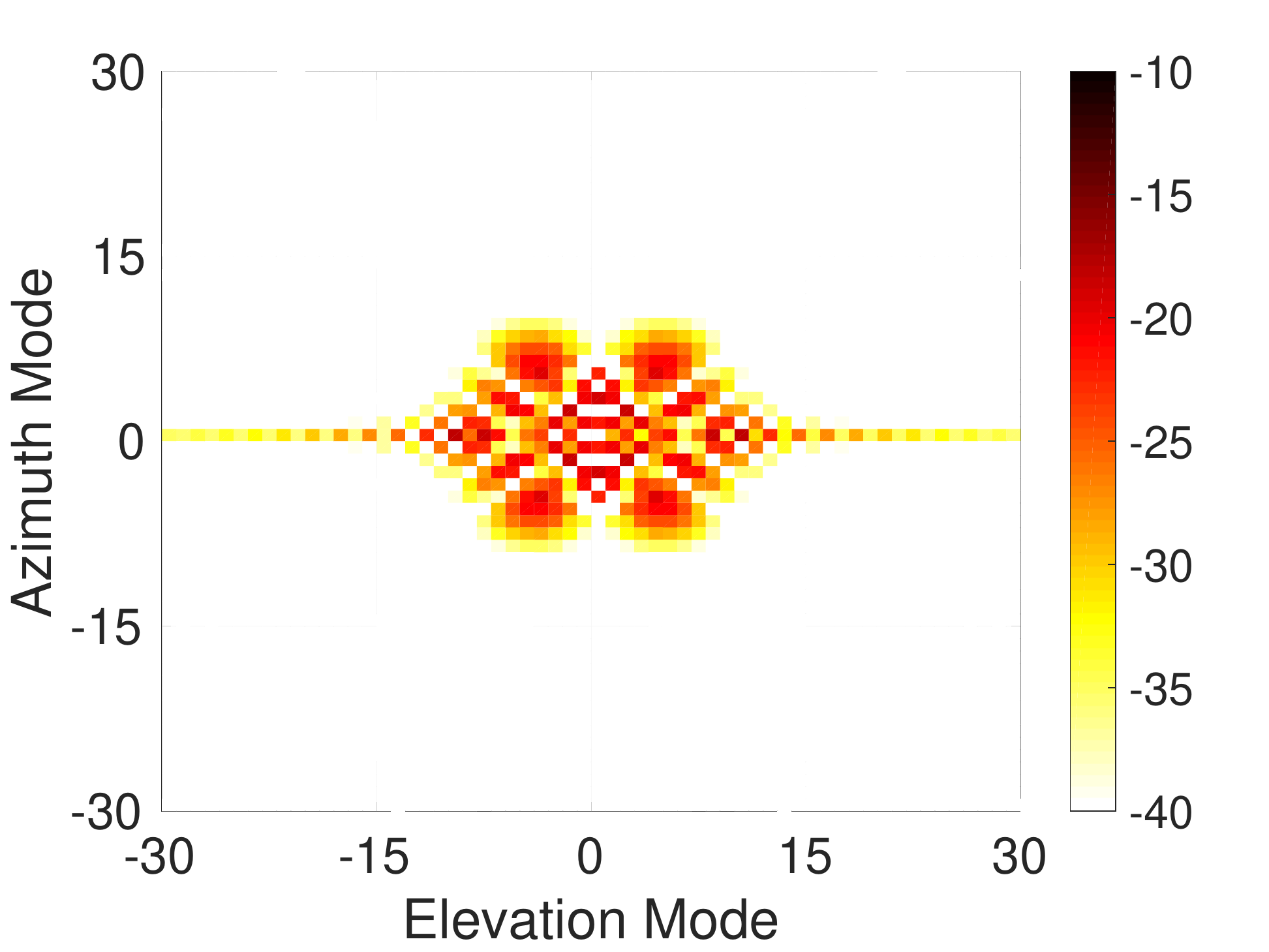}} \\
    \subfigure[$\Delta\B{p}=2\lambda\B{1}$]{\includegraphics[width = 0.4\columnwidth, 
  clip = true]{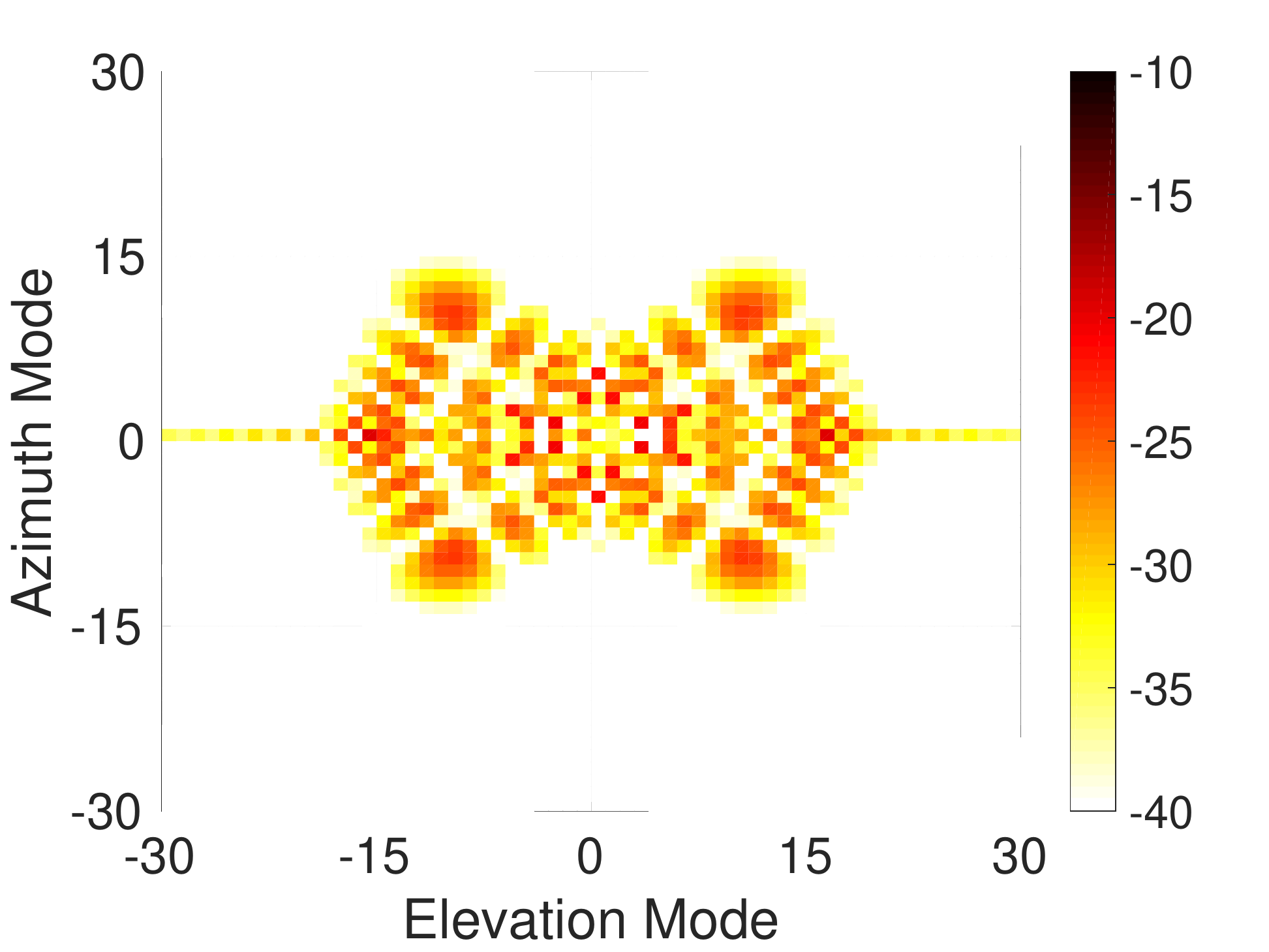}}
  \subfigure[$\Delta\B{p}=3\lambda\B{1}$]{\includegraphics[width = 0.4\columnwidth, 
  clip = true]{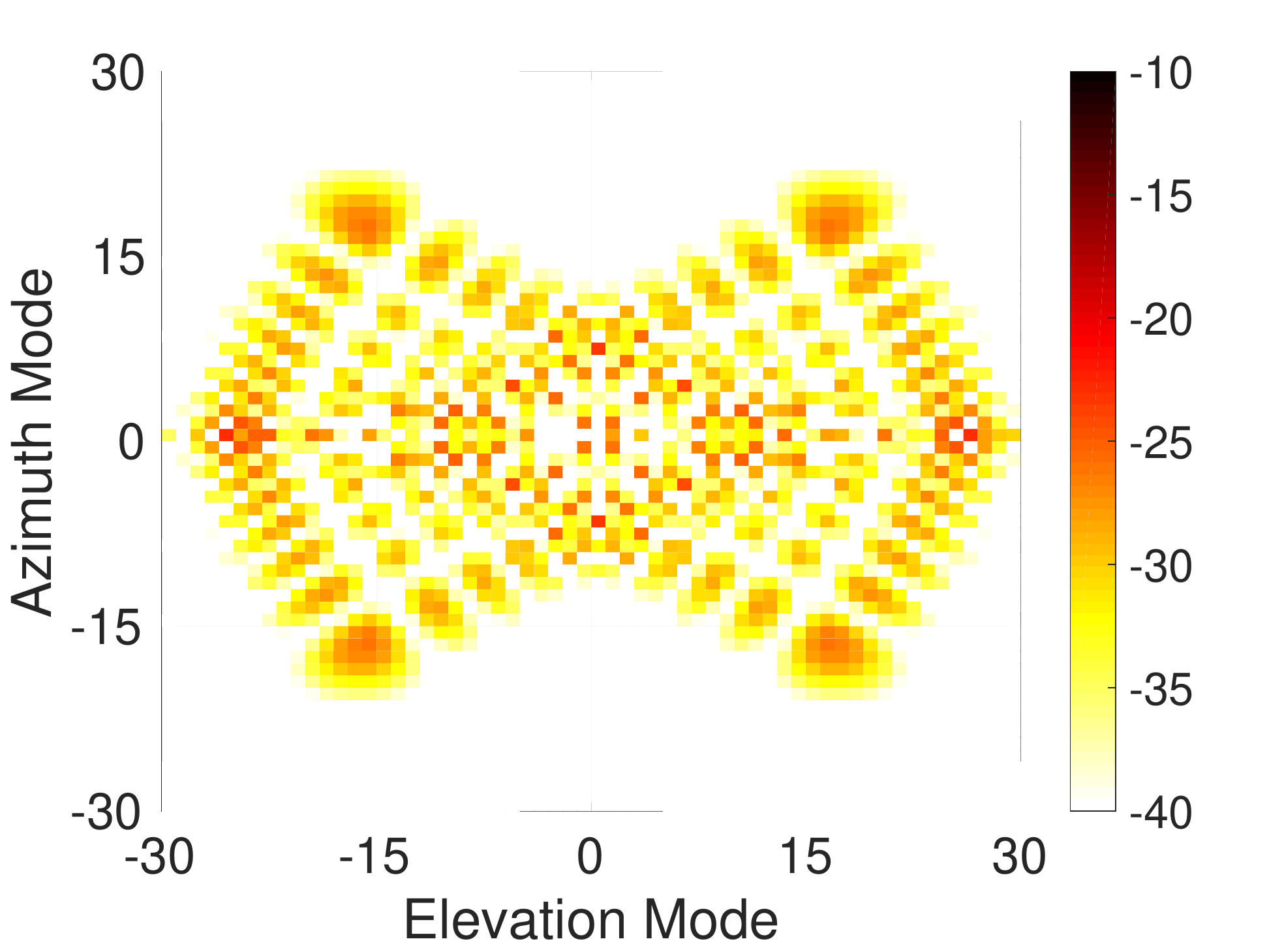}\label{subfig:EADF_3lambda}} 
  \caption{Amplitude of \gls{eadf} of ($n_y=1,n_z=1$) in the $8 \times 2$ \gls{ura} with different center offset values $\Delta\B{p}$} 
  \label{fig:EADF_comp_misalign}
\end{figure}

The array ambiguity function is used to check the performance of the array to differentiate signals from different directions \cite{wang2018channel}. It is usually defined as 
\begin{equation}
  A_b(\varphi_1,\theta_1,\varphi_2,\theta_2) = \frac{\B{b}^\dagger(\varphi_1,\theta_1)\B{b}(\varphi_2,\theta_2)}{\lVert\B{b}^\dagger(\varphi_1,\theta_1)\rVert \cdot \lVert\B{b}(\varphi_2,\theta_2)\rVert}.
\end{equation}
We can replace the second $\B{b}$ with $\tilde{\B{b}}$ in the above equation and examine the ``cross'' ambiguity function. Fig. \ref{fig:XAmbFunc} shows that it presents a high ridge in the off-diagonal direction, which means that if the actual response is $\B{b}$ while the calibrated array response is $\tilde{\B{b}}$, the estimator is very likely to provide the correct result, as shown through the following simulation results.

\begin{figure}[!t]
  \centering
  \includegraphics[width = \wid\columnwidth]{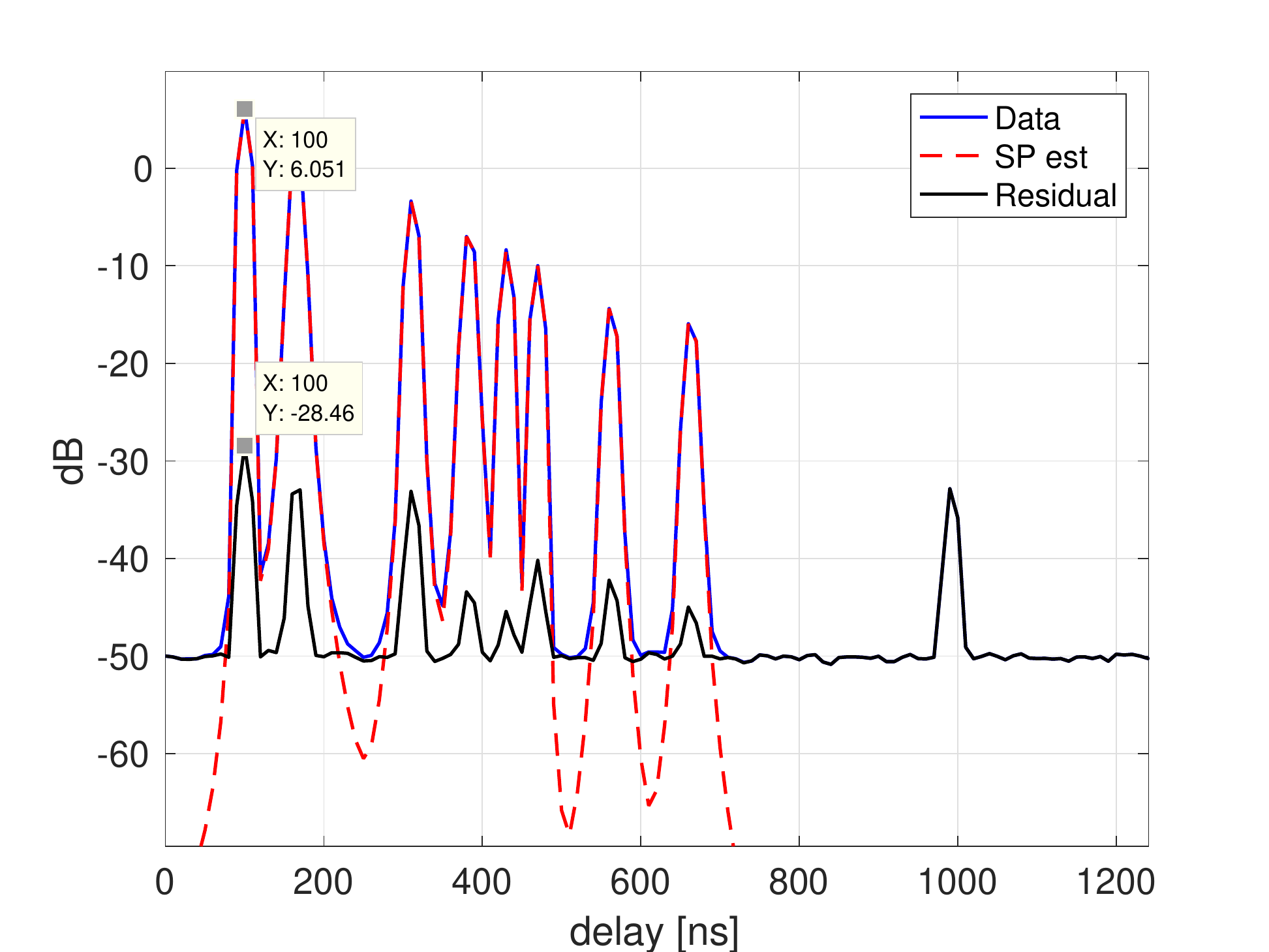}
  \caption{The \gls{apdp} comparison between the synthetic channel response, the reconstructed channel response from Rimax estimates and the residual channel response based on center-misaligned RFU patterns}
  \label{fig:APDP_off3lambda_sim}
\end{figure}

We provide simulation results with Rimax evaluation based on synthetic channel responses, in order to study the impact of center misalignment and phase noise during the calibration. A similar approach to generate synthetic channel responses is also presented in Ref. \cite{wang2015efficiency}. 
The carrier frequency is set to \SI{28}{GHz} and the bandwidth is \SI{100}{MHz}. Both the \gls{tx} and \gls{rx} arrays follow the \gls{ura} configuration shown in Fig. \ref{fig:URA_2D}.
We have simulated channel responses when the \textit{ideal} calibrated array response $\B{b}$ is used. In the Rimax evaluation, we then use the \glspl{eadf} extracted from the ``distorted'' patterns when the center offset $\Delta\B{p} = 3\lambda\B{1}$, {\color{black}which is the upper limit on the center misalignment considering our efforts to align the probe with the RFU}. Fig. \ref{subfig:EADF_3lambda} provides the \gls{eadf} amplitude pattern for one of the corner elements. In Fig. \ref{fig:APDP_off3lambda_sim} we compare the \glspl{apdp} of the synthetic channel response, the reconstructed channel response based on Rimax estimates and the residual channel response due to the center misalignment. The peak reduction is around \SI{30}{dB} for each path. {\color{black} We define the peak reduction as the power difference between \gls{apdp} peaks of the original signal and the residual signal after the parameter estimation.} Tab. \ref{Tab:ParameterComp} provides the comparison of parameters for each path. Because of the imperfect amplitude estimate of path 1, the residual peak of path 1 is still higher than that of path 10 in this simulation, and consequently the estimator fails to detect path 10. 

\begin{table*}[!t]
  \small
  \centering
  \caption{Compare the estimated path parameters with the center misaligned array responses, format true/estimated}
  \label{Tab:ParameterComp}
  \begin{tabular}{c|cccccc}
     \toprule
    Path ID & $\tau$(ns) & $\varphi_T$ (deg) & $\theta_T$ (deg) & $\varphi_R$ (deg) & $\theta_R$ (deg) & $P$ (dBm) \\
     \midrule
     1 & 100.00/100.00 & 18.45/18.67 & 109.53/108.60 & -31.21/-30.70 & 104.56/104.36 & -20.00/-20.00\\
     2 & 165.17/165.17 & -0.70/-0.33 & 90.00/89.51 & 46.38/45.97 & 53.39/53.91 & -22.83/ -22.85\\
     3 & 311.45/311.45 & 33.49/33.60 & 88.39/87.37 & -56.56/-55.77 & 55.72/55.47& -29.18/-29.20 \\
     4 & 383.01/382.99 & 25.80/26.07 & 122.38/121.98 & -1.21/-0.84 & 91.73/91.36 & -32.29/-32.31 \\
     5 & 430.16/430.16 & 48.45/48.48 & 98.79/98.66 & -39.85/-39.14 & 57.74/57.64 & -34.34/-34.35 \\
     6 & 468.86/468.86 & 46.91/46.89 & 99.41/98.95 & 57.44/56.81 & 115.45/114.16 & -36.02/-36.02 \\
     7 & 561.61/561.60 & -19.90/-19.43 & 118.76/118.68 & 25.52/25.53 & 115.40/114.16 & -40.05/-40.04 \\
     8 & 661.73/661.74 & 23.85/23.97 & 114.44/113.21 & 0.06/0.47 & 107.80/105.80 & -44.40/-44.39\\
     9 & 662.94/662.94 & -36.26/-35.76 & 96.14/95.58 & -3.47/-3.03 & 61.99/61.88 & -44.45/-44.52 \\
    $10^\ast$ & 990.65/100.00 & -56.34/3.59 & 64.63/-19.28 & -52.85/-30.48 & 102.77/102.41 & -58.68/-52.66 \\
    \bottomrule
  \end{tabular} 
\end{table*}

\subsection{Phase Noise}
\label{sect:PN_model}
\begin{figure}[!t]
  \centering
  \includegraphics[width=\wid\columnwidth,clip=true]{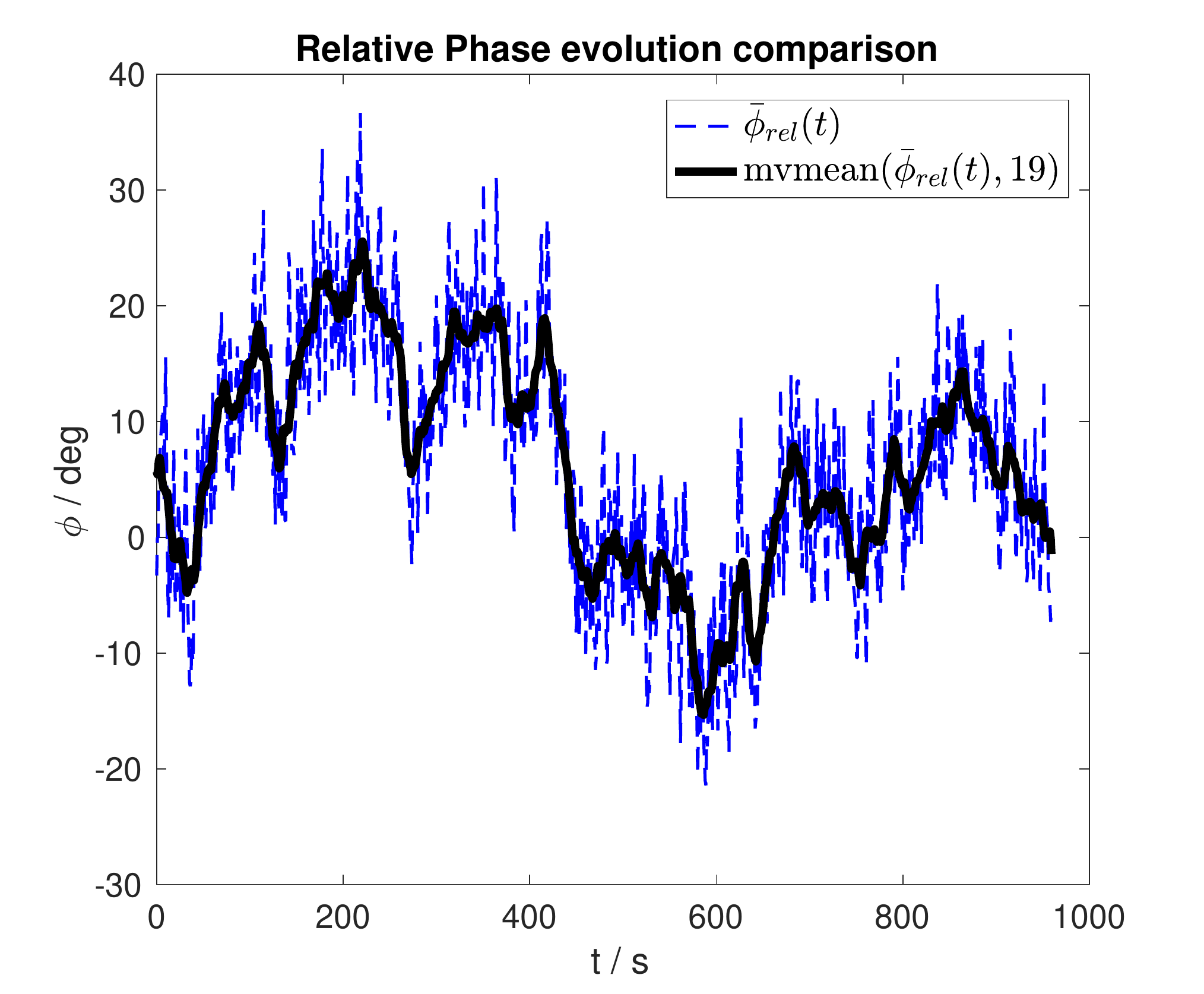}
  \caption{The relative phase response of two RFUs measured in an anechoic chamber over a 15-min timespan}
  \label{fig:phase_test}
\end{figure}

{\color{black}Because the \gls{lo} signals in two RFUs are generated separately, although the \gls{tx} and \gls{rx} RFUs share the same \SI{10}{MHz} reference clock, there still remain some small phase variations.} 
To study the potential impact of phase noise on the pattern calibration in Section \ref{sec:calibProc}, we measured the system phase response for about 15 minutes. We denote the time-varying S21 measurement as $S_{21}(f,t)$, and the normalized phase response as $\phi_\tx{rel}(f,t)=\tx{arg}(S_{21}(f,t)/S_{21}(f,0))$\footnote{The arg($z$) function returns the phase in radians of the complex variable $z$}. As shown in Fig. \ref{fig:phase_test}, the relative phase averaged over frequency $\bar{\phi}_\tx{rel}(t)=\mathbb{E}_f\{\phi_\tx{rel}(f,t)\}$ shows a \textit{combination} of fast and slow variations over time. 

\begin{figure}[!t]
  \centering
  \subfigure[The empirical autocorrelation]{\includegraphics[width=0.45\columnwidth,clip=true]{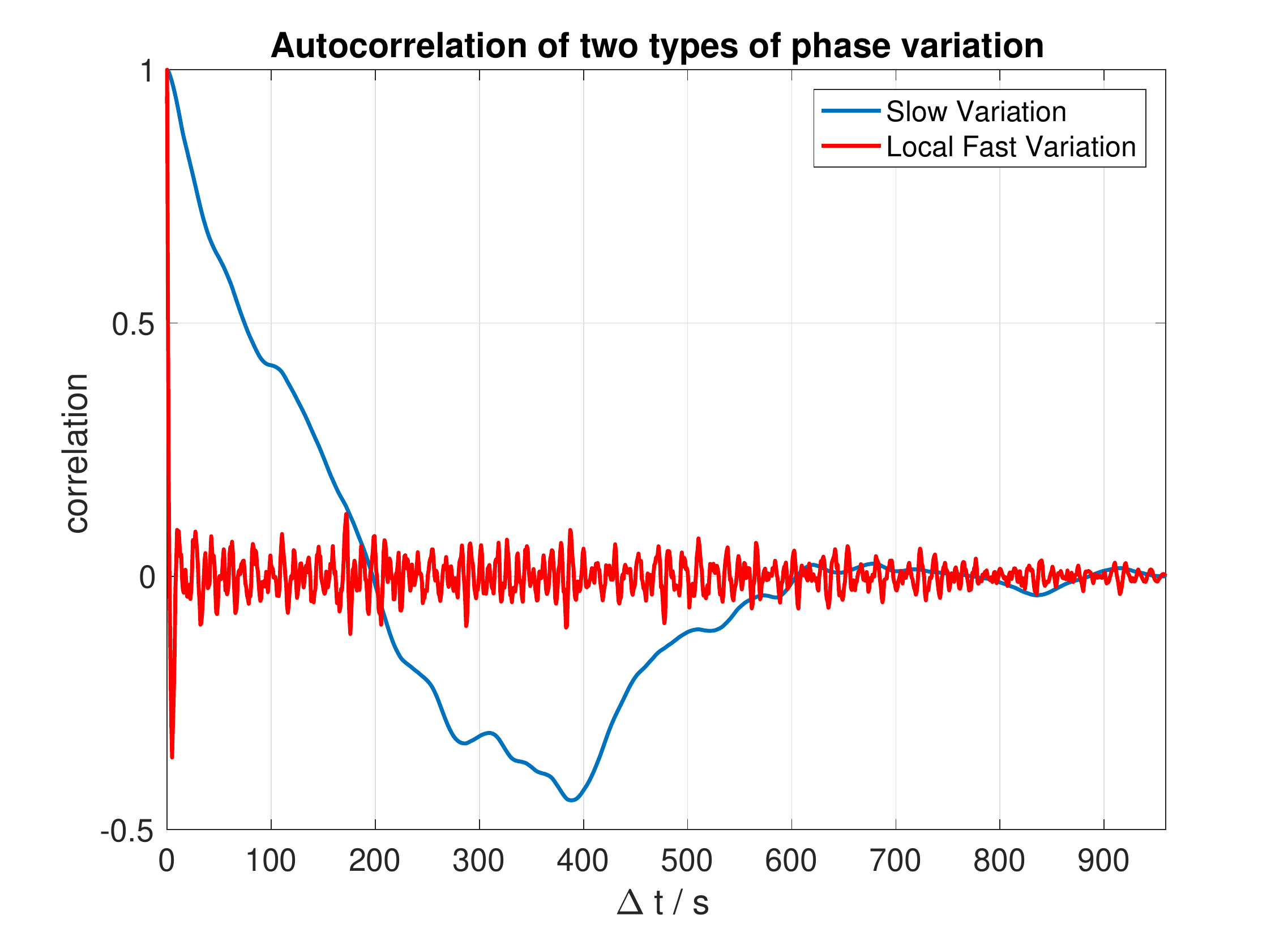}
  \label{fig:phase_resp_correlation}}
  \subfigure[The \gls{cdf} of the fast variation]{\includegraphics[width=0.45\columnwidth,clip=true]{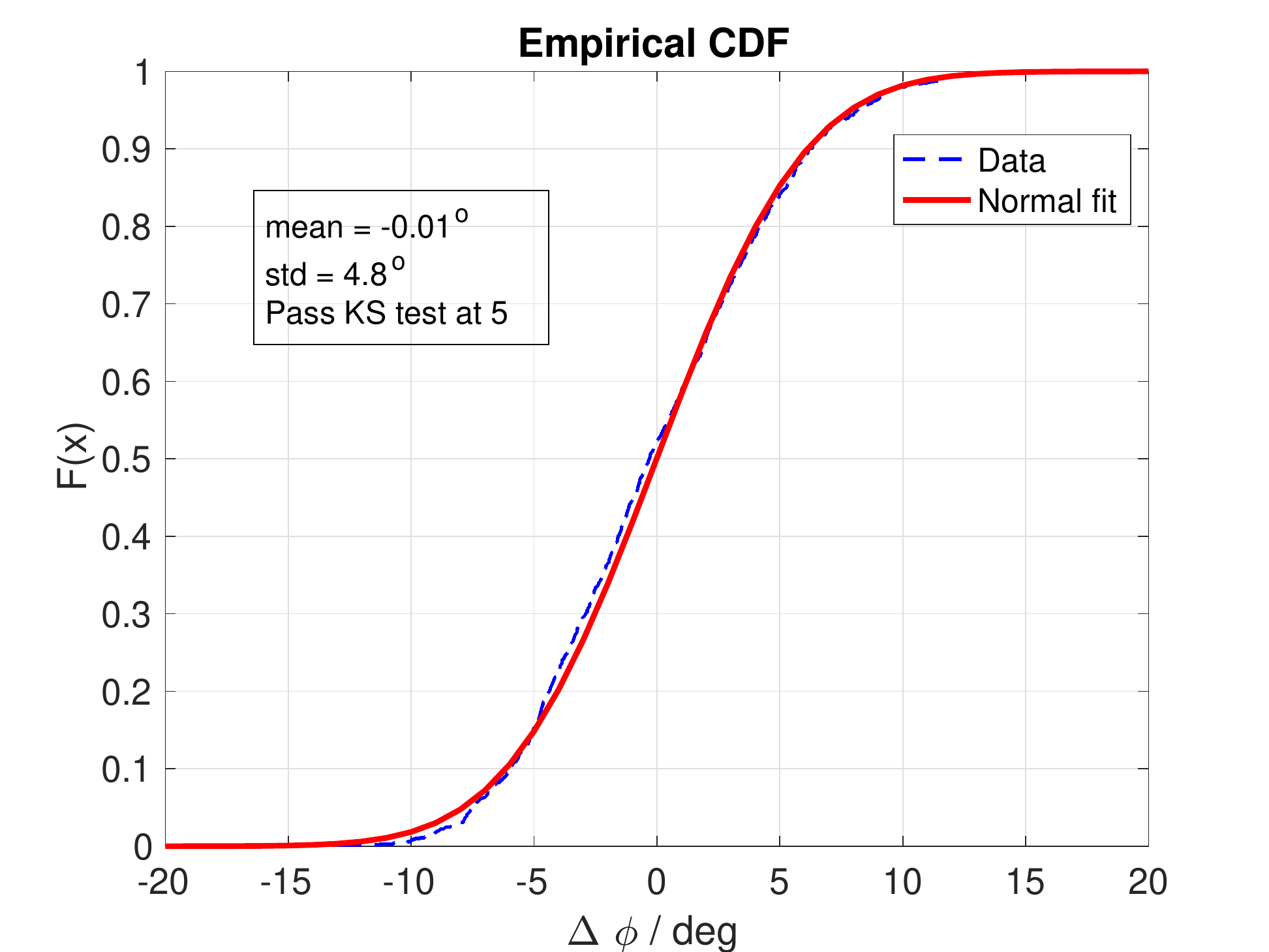}
  \label{fig:local_phase_stats}}
  \caption{The statistics of the ``slow'' and ``fast'' phase variations based on the phase stability test} 
\end{figure}

The slow-varying phase response can be observed by performing a moving average on $\bar{\phi}_\tx{rel}(t)$, and Fig. \ref{fig:phase_resp_correlation} presents the autocorrelations of two types of variations. The local fast phase variation can be closely modeled by an i.i.d. Gaussian process with a mean $-0.01^\circ$ and a standard deviation $4.8^\circ$. Its fitting to a normal distribution passes the two-sample KS test with a $5\%$ confidence level. 

%
To model its influence on the measured response, we consider the calibration scheme highlighted in Section \ref{sec:calibProc} and assume that there is an i.i.d. Gaussian variation between switched beams at the same panel rotation angle, while another slow-varying term is kept constant within different beams at the same angle and only changes between orientations. 

\begin{figure}[!t]
  \centering
  \includegraphics[width=\wid\columnwidth,clip=true]{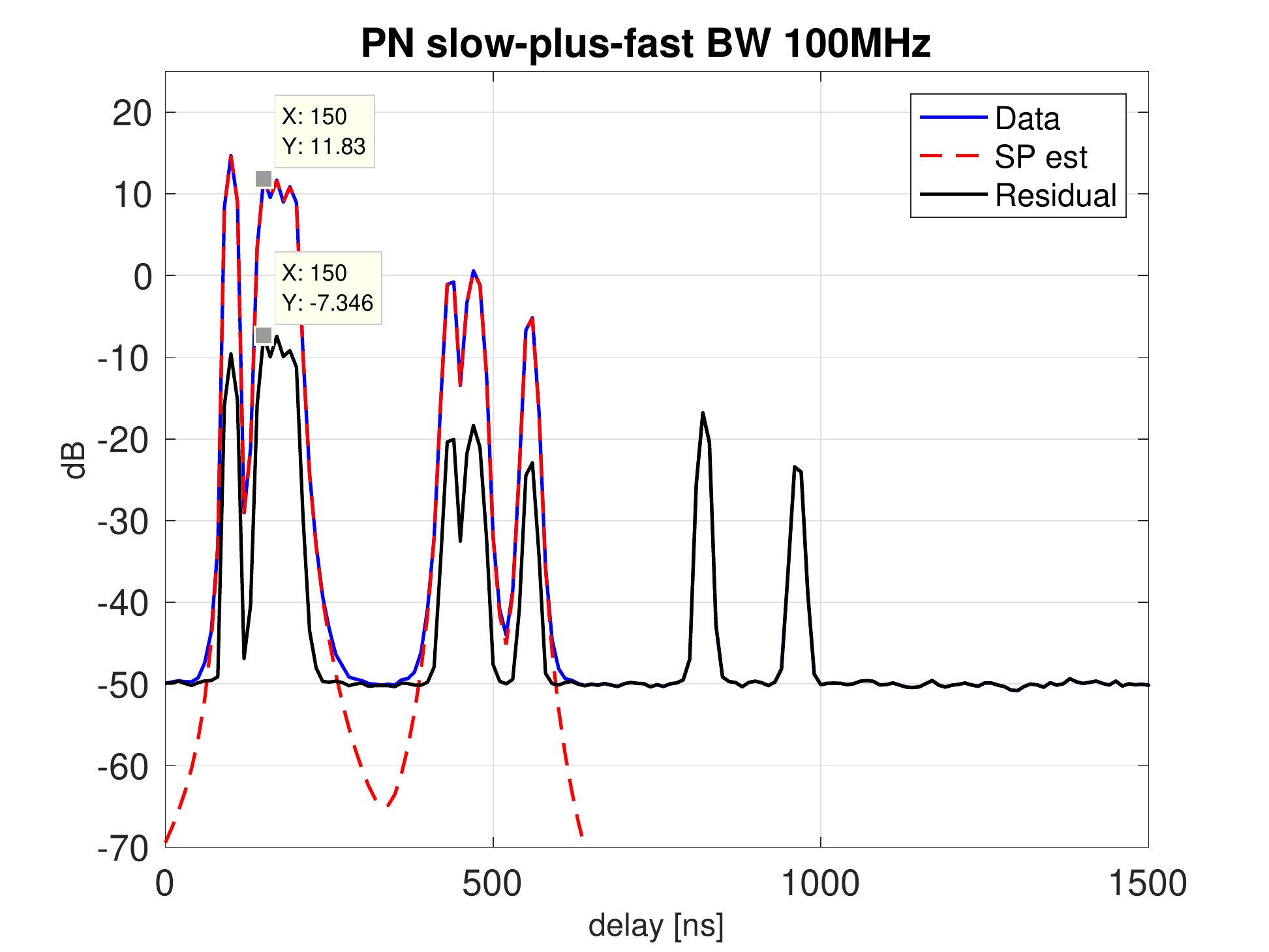}
  \caption{The \gls{apdp} comparison between the synthetic channel response, the reconstructed channel response from Rimax estimates and the residual channel response based on PN-corrupted RFU patterns}
  \label{fig:APDP_PN_sim}
\end{figure}

We then conduct simulations to study the effects of residual phase noise in the calibrated pattern on Rimax evaluation. 
Phase noise also impacts the actual measurement data obtained in real-time, however they are only corrupted with the fast phase variation but free from the slow-varying \gls{pn}, because each \gls{mimo} measurement with the time domain setup indicated in Fig. \ref{fig:summary_diagram}(a) takes about \SI{1.4}{ms}, which is insignificant when compared with the correlation time of the slow-varying phase term shown in Fig. \ref{fig:phase_resp_correlation}.
Fig. \ref{fig:APDP_PN_sim} shows the comparison of \glspl{apdp} after we perform a Rimax evaluation with the PN-corrupted RFU patterns. The peak reduction is only \SI{18}{dB} in this scenario.


In summary, the limiting factor on the performance of Rimax evaluation with our mmWave channel sounder is the \gls{pn} corruption on the calibrated RFU patterns, where the peak reduction is limited to about \SI{20}{dB} for each path, and it could lead to the existence of ``ghost'' paths as well as failures to detect relatively weaker paths at larger delays. However it may be sufficient in most of \textit{practical} environments, such as indoor office and outdoor macrocell, where the power of \gls{dmc} could contribute between $10\%$ and $90\%$ of the total power \cite{richter2006distributed,poutanen2011angular}, therefore the residual signals would be below the spectrum of \gls{dmc} and have a limited influence on the estimation results especially when the Rimax-based estimator considers \gls{dmc} as a part of the channel response \cite{richter2005joint}. 

\section{Measurement}
\label{sec:Meas}
{\color{black} In this section, we first summarize the steps to apply the calibration results from Section \ref{sec:calibProc} to pre-process the measurement data before running an \gls{hrpe} algorithm. Secondly, we conduct \gls{mimo} channel measurements in an anechoic chamber and present the Rimax evaluation results for two types of test channels. The first is the two-path channel that consists of a direct \gls{los} signal and a strong reflection from a metallic plane. The second is the two-pole test channel where two standing poles provide potential weaker reflections, and we also move one of the poles to create different angular separation between the two reflections.}

\subsection{Preprocessing with Calibration Results}
We provide the recipe to apply various calibrated frequency responses to the measurement data before putting them into Rimax. The measurement data are generally produced by the time-domain setup given in Fig. \ref{fig:summary_diagram}(a), and we denote them as $Y_\tx{data}(f,m,n)$. We calibrate the through cable between \SI{1.65}{GHz} and \SI{2.05}{GHz}, represented by the green line in Fig. \ref{fig:summary_diagram}(b), with a \gls{vna} and obtain its response $H_\tx{cable}(f)$. On the other hand, the data generated by the through calibration in Fig. \ref{fig:summary_diagram}(b) is denoted as $Y_\tx{IF}(f)$. The common frequency response of two RFUs, which is produced by the two-step calibration and extracted according to Alg. \ref{Alg:Jt_EADFExt_vec} and Alg. \ref{Alg:MultiGain_EstVerify}, is denoted as $G_0(f)$.

As a result, the frequency compensation to the original data $Y_\tx{data}$ is given by
\begin{equation}
  \tilde{Y}(f,m,n) = \frac{Y_\tx{data}(f,m,n)}{G_0(f)Y_\tx{IF}(f)/H_\tx{cable}(f)},
\end{equation}
which is ready as an input to a super-resolution estimation algorithm such as Rimax or \gls{sage}.

\subsection{Two-path Experiment}
\begin{figure}[!t]
  \centering 
  \subfigure[A picture]{\includegraphics[width=0.4\columnwidth,clip=true]{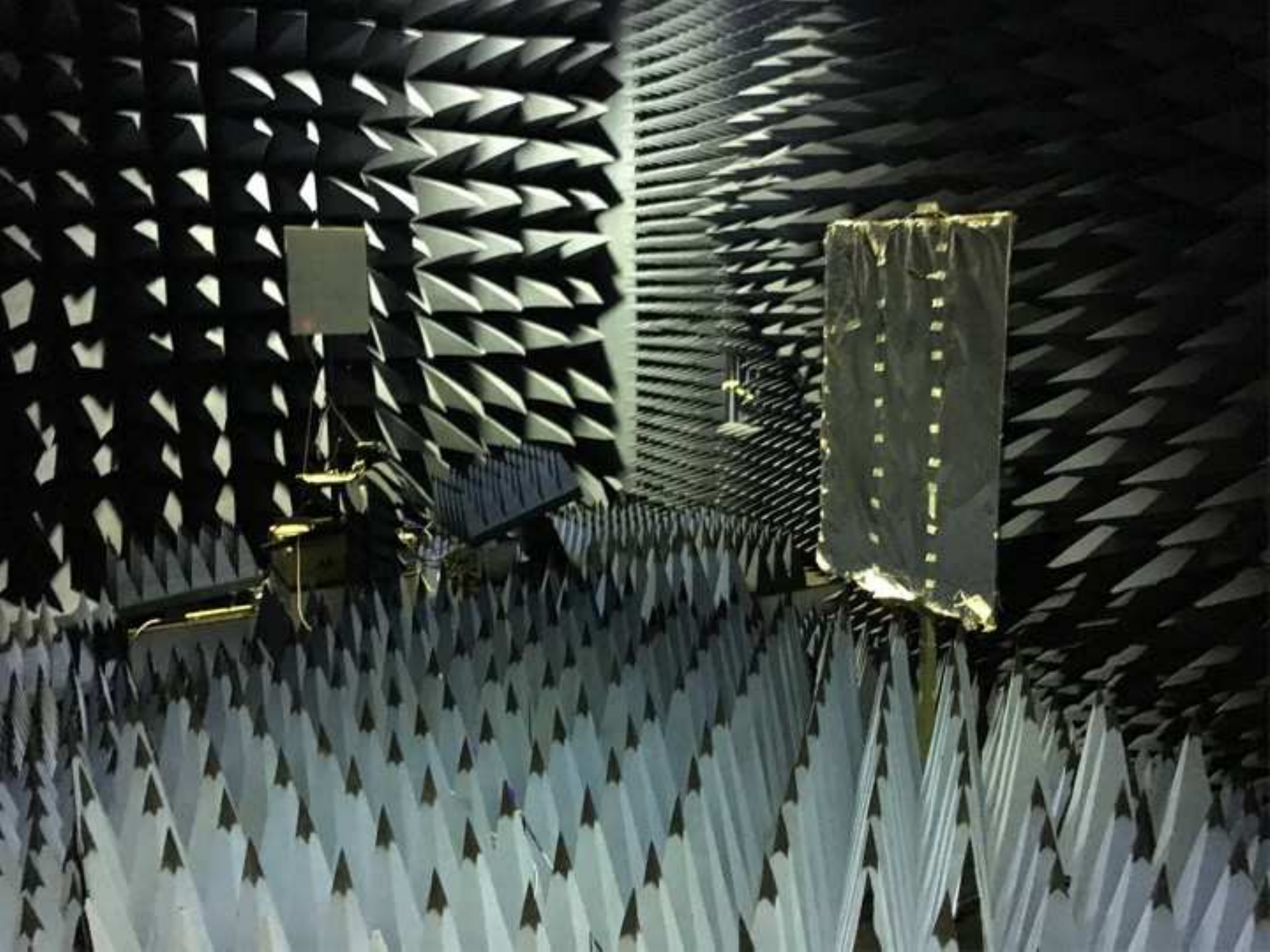} \label{fig:chamber_verify_picture}}
  \subfigure[The diagram with birdview]{\includegraphics[width=0.45\columnwidth,clip=true]{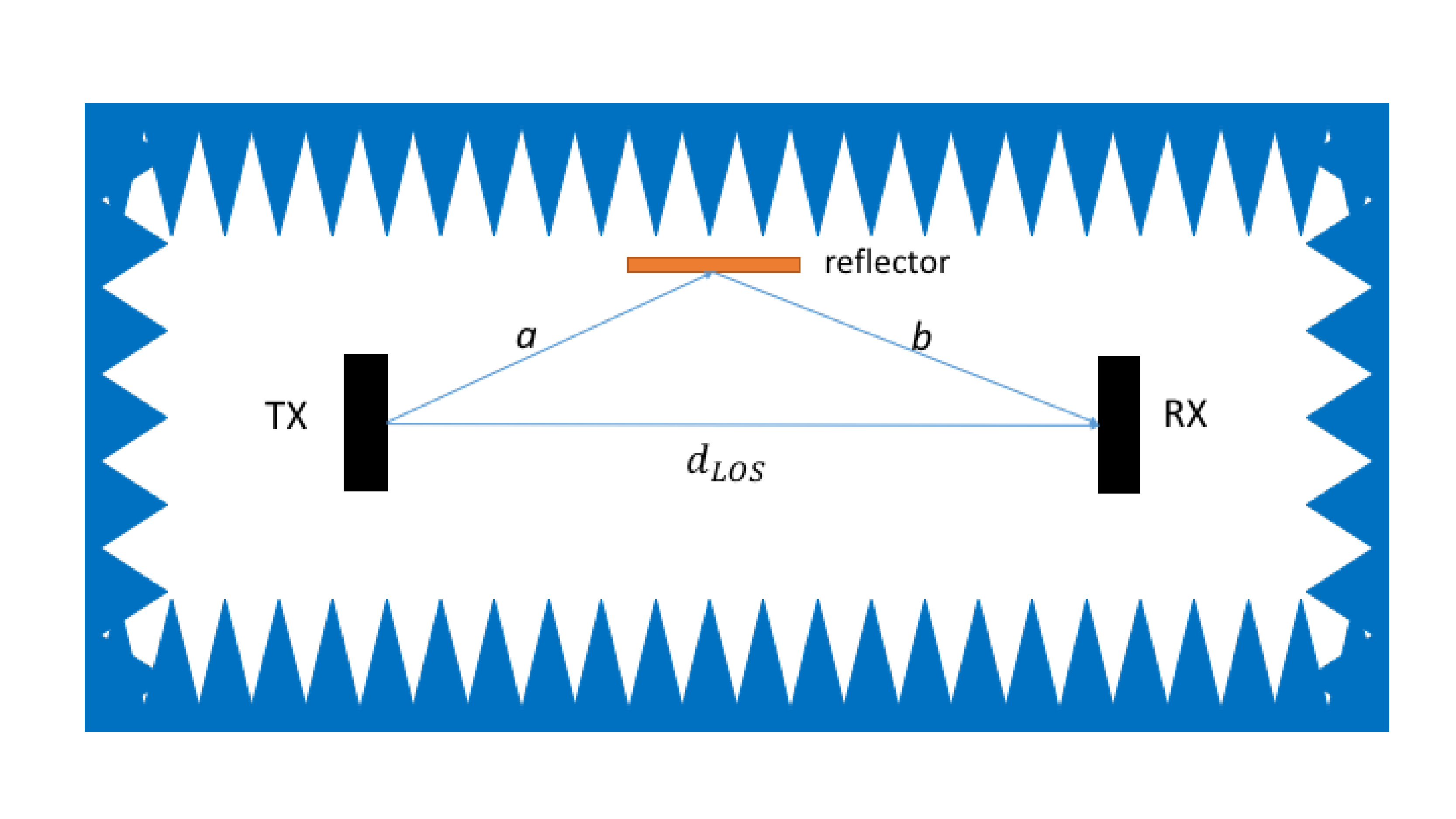}
  \label{fig:chamber_verify_diagram}}
  \caption{The two-path test channel with mmWave RFUs in the anechoic chamber}
\end{figure}

To experimentally verify the calibration results, we created a two-path test channel in an anechoic chamber with our \gls{mimo} \gls{mmwave} channel sounder \cite{bas2017real}. The setup is illustrated in Fig. \ref{fig:chamber_verify_picture} and \ref{fig:chamber_verify_diagram}. Although the reflector is a plane, since we enforce the reflector to be parallel to the wall of the chamber, the actual specular reflection \textit{point} should have the same distance to both \gls{tx} and \gls{rx}. Consequently the distances indicated in Fig. \ref{fig:chamber_verify_diagram} are $a=b=\SI{3.15}{m}$, and $d_\tx{LOS}=\SI{5.65}{m}$. Based on trigonometry, the azimuth \gls{dod} of the reflected path is around $26^\circ$ and the azimuth \gls{doa} is about $-26^\circ$. The extra path delay of the reflection, when compared with \gls{los}, is about \SI{2.17}{ns}, {\color{black}which is about one quarter of the inverse of the \SI{100}{MHz} bandwidth}. The channel is basically composed of one dominant LOS path and a reflection from a metallic plane. The metallic plane is elevated to the same height of \gls{tx} and \gls{rx} RFUs, hence we can assume all significant signals travel in the azimuth plane.   

We analyzed in total 470 \gls{mimo} snapshots with Rimax following the steps in Section. \ref{sec:calibProc}. The statistics, such as mean and standard deviation, of path parameters is listed in Tab. \ref{Tab:two_path_chamber}.  Path 1 is the \gls{los} path, and Path 2 is the reflected path, and both of their path directions match well with the geometry. The theoretical \gls{los} power is \SI{-26.03}{dB}, which is calculated by $1/4\pi d^2$. The extra delay of Path 2 is \SI{2.25}{ns}, which equals \SI{67.5}{cm} in path length and is close to the \SI{65}{cm} computed based on the geometry. Path 3 is the double-reflection between two RFUs when they are aligned and facing each other. The extra run length is about \SI{10.55}{m} which is close to two times of $d_\tx{LOS}$. The peak reduction is about \SI{12}{dB} and the \textit{unresolved} residual signal power is less than 8\% of the original signal, which is somewhat close to the peak reduction level in the \gls{pn}-corrupted-data simulation. Path 4 is a ``ghost'' path as its power is more than 10 dB lower than that of Path 2 and it cannot be physically mapped to any reflector in the chamber. Based on the results of this verification measurement and the simulations in Section \ref{sec:calibLimit}, paths with similar delay values (less than \SI{1}{ns}) whose power are more than \SI{10}{dB} weaker than the strongest one should be considered as ``ghost'' paths and discarded for further processing. 

\begin{table}[!t]
  \centering
  \caption{List of Rimax estimates for the the two-path test channel in the anechoic chamber, mean / standard deviation}
  \label{Tab:two_path_chamber}
  \begin{tabular}{c|cccc}
     \toprule 
      Path ID & $\tau$ (ns) & $\varphi_T$ (deg) & $\varphi_R$ (deg) & $P$ (dB) \\
     \midrule
      1     & 20.30/0.06 & -1.43/0.06 & -0.14/0.03 & -23.80/0.06 \\
      2     & 22.55/0.07 & 26.27/0.07 & -25.55/0.19 & -28.83/0.23 \\
      3     & 55.47/0.08 & -1.18/0.07 & -0.25/0.02  & -40.94/0.11 \\
      4     & 22.40/7.58 & 6.54/17.30 & 13.57/30.92 & -46.74/3.36 \\ 
     \bottomrule
  \end{tabular}
\end{table}


\begin{figure}[!t]
  \centering
  \includegraphics[width=\wid\columnwidth,clip=true]{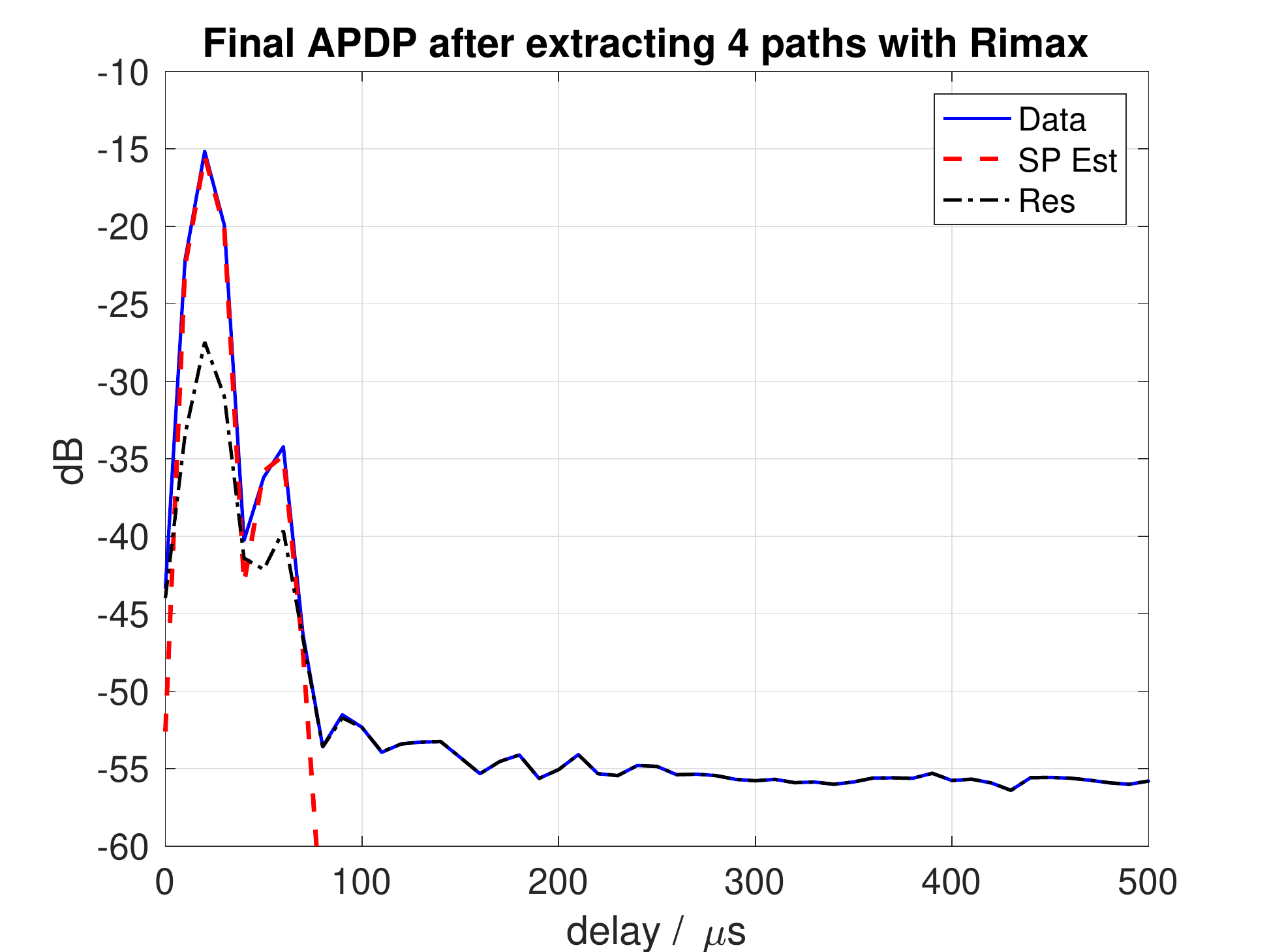}
  \caption{The APDP comparison between the two-path test channel, the reconstructed two-path test channel and the residual response}
\end{figure}

\subsection{{\color{black}Two-pole Experiment}}
\begin{figure}[!t]
  \centering
  \includegraphics[width=\wid\columnwidth,clip=true]{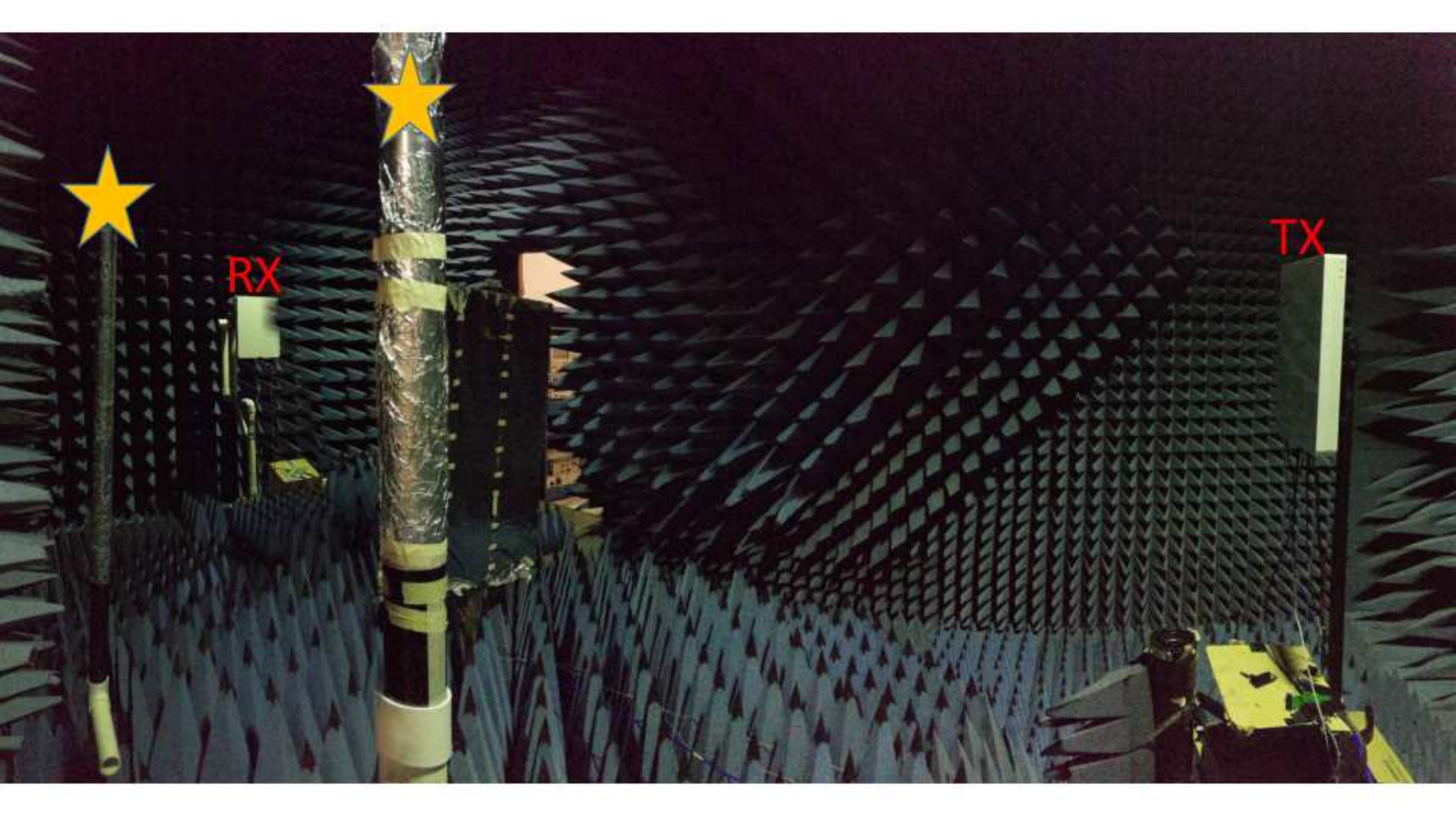}
  \caption{The picture of the two-pole experiment in an anechoic chamber, the two poles are marked with yellow stars.}
  \label{fig:two_pol_exp_setup}
\end{figure}

We further investigate the capability of the setup to differentiate signals with a limited angular separation, and compare the results with the geometry as well as the Fourier-based beamforming \gls{aps}. The measurement setup can be found in Fig. \ref{fig:two_pol_exp_setup}. We place two standing posts with their upper parts wrapped with reflective aluminum foil. The diameters of these posts are small so that they are weaker reflectors compared to the metallic plane in the previous two-path experiment. The metallic plane is inserted between \gls{tx} and \gls{rx} so that the direct \gls{los} is attenuated. 

\begin{figure}[!t]
  \centering 
  \subfigure[Position 1]{\includegraphics[width = 0.45\columnwidth, 
  clip = true]{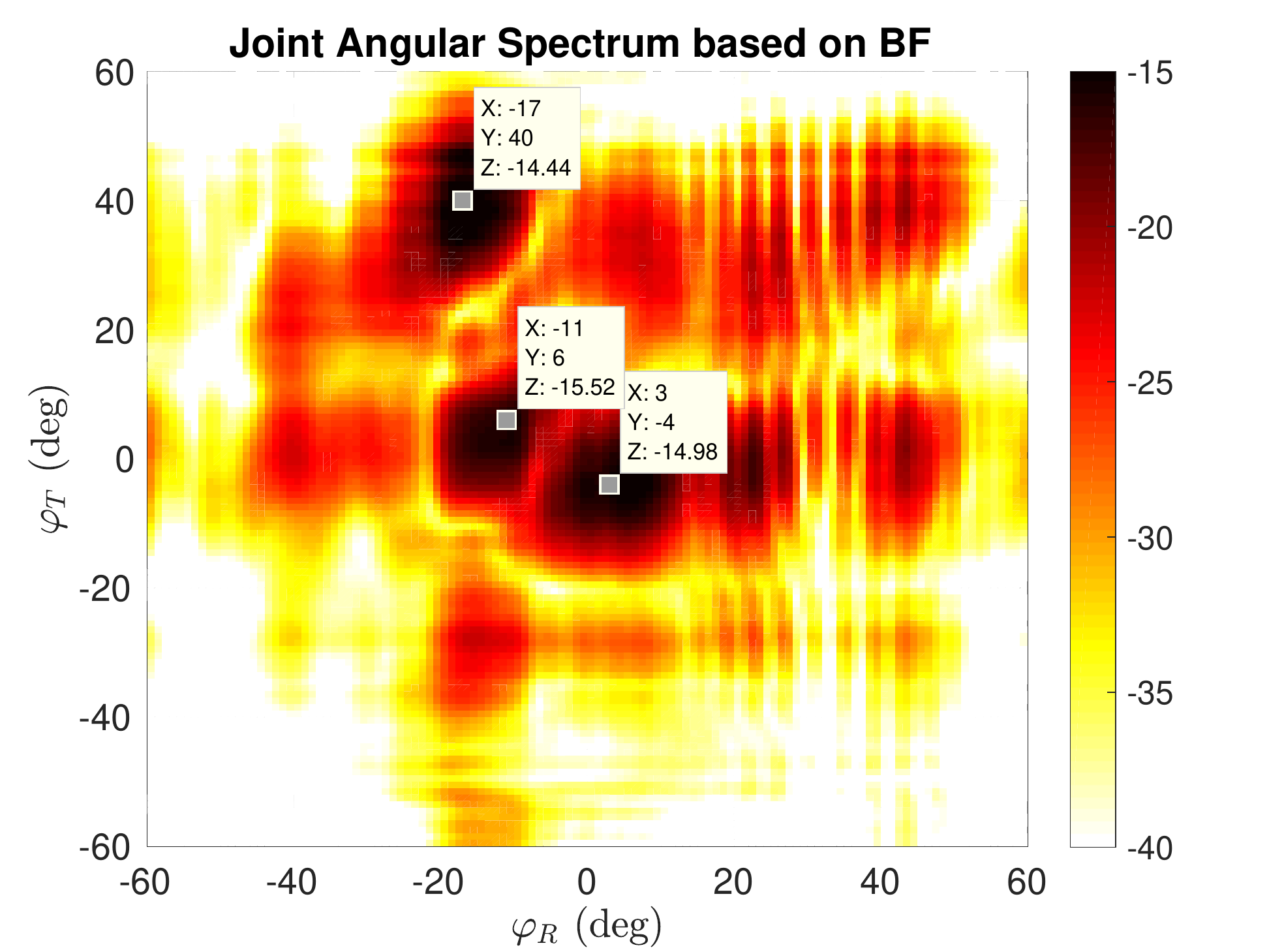}\label{subfig:two_pol_2Dbf_pos1}} 
  \subfigure[Position 4]{\includegraphics[width = 0.45\columnwidth, 
  clip = true]{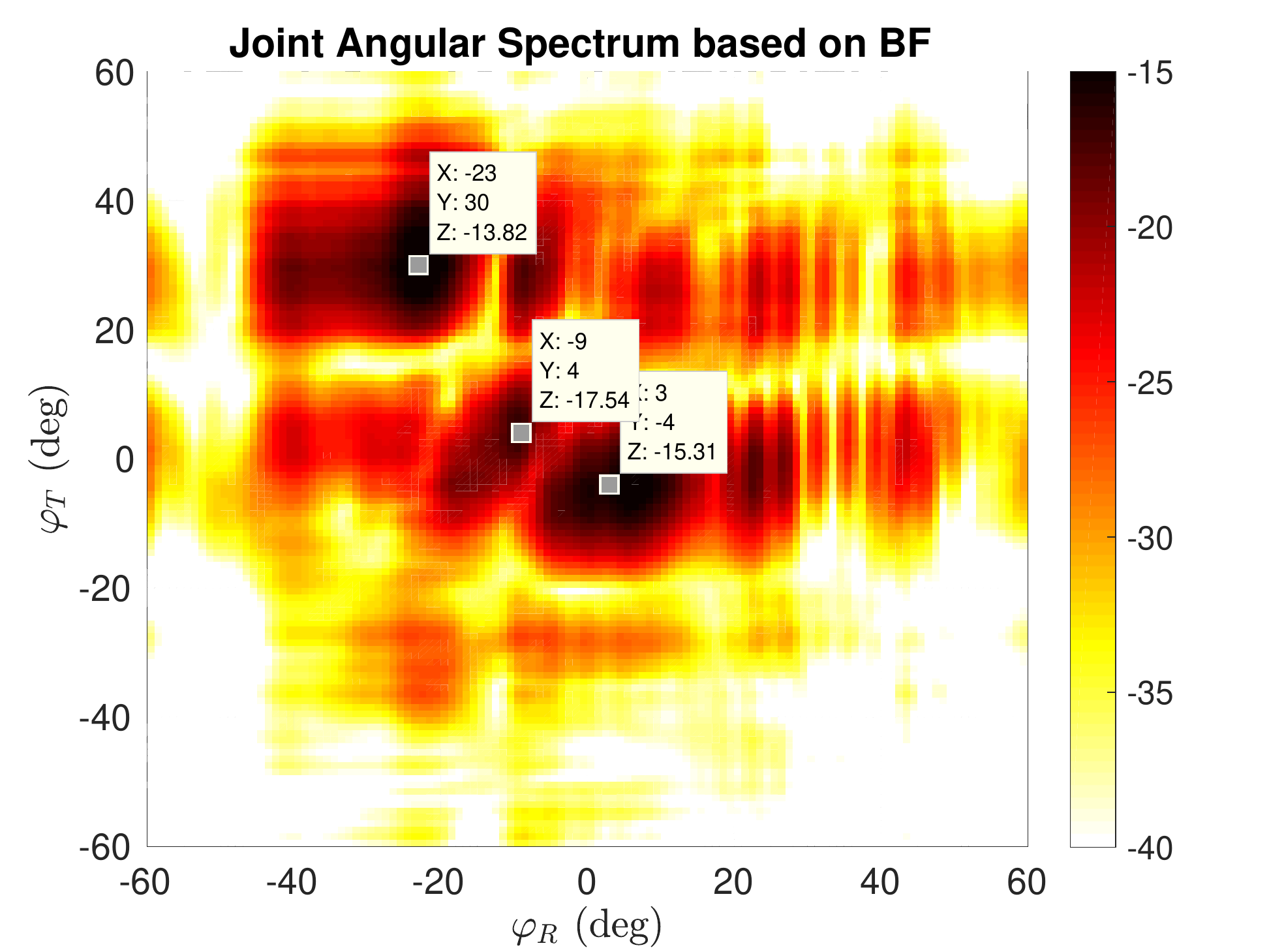}\label{subfig:two_pol_2Dbf_pos4}} 
  \caption{The two-dimensional beamforming \gls{aps} of the measurements for position 1 and position 4 of the two-pole measurement}
  \label{fig:bf_two_pol}
\end{figure}

We name the post that is closer to the \gls{tx} RFU in Fig. \ref{fig:two_pol_exp_setup} as post A, and the other as post B. We  move post A to five different positions so that the angular separation of reflected signals from the two posts gradually decreases. For one of the positions, a complete parameter list of six paths is provided in Tab. \ref{Tab:two_pol_results_pos1}. Path 2 and 5 are from post A and B, respectively. The angular difference is about $17^\circ$ for the azimuth \gls{dod} and $10^\circ$ for the azimuth \gls{doa}. The estimates of two diffraction signals around the metallic plane are consistent when post A moves among the five locations. The summary of the estimated reflections from post A and post B is available in Tab. \ref{Tab:two_pol_twoReflect}. We can observe that the estimated azimuth \glspl{dod} of \textit{both} reflections and the diffracted signals have an approximate offset of $5^\circ$ when compared to their true values based on the geometry. This offset is most likely due to the mismatch of orientation angles of RFUs in the setup. However the estimated angular \textit{differences} are aligned with those from the true values, and thus the estimated angular spreads will also be close. Besides, we also see that the calibrated setup with Rimax is unable to see two reflected signals from position 4, where the azimuth \gls{doa} separation is about $5^\circ$. This angular separation limit is certainly affected by the imperfect estimation of other strong signals such as the diffractions and other reasons mentioned in Section \ref{sec:calibLimit}. If we compare these results with the Fourier-based beamforming results of position 1 and position 4 in Fig. \ref{fig:bf_two_pol}, we could see in Fig. \ref{subfig:two_pol_2Dbf_pos1} that even for position 1 when two posts have the largest separation, the \gls{aps} appears to fail to produce two distinguishable peaks around the region that matches the directions of the posts, thus showing the clear superiority of HRPE. 

\begin{table}[!t]
  \centering
  \caption{List of Rimax estimates for the the two pole measurement in an anechoic chamber when post A is at position 1}
  \label{Tab:two_pol_results_pos1}
  \begin{tabular}{c|cccc}
    \toprule 
      Path type & $\tau$ (ns) & $\varphi_T$ (deg) & $\varphi_R$ (deg) & $P$ (dB) \\
    \midrule
      Diffraction & 18.81 & -4.62 & 3.32 & -37.99 \\
      Post A      & 20.68 & 40.10 & -17.44 & -40.29 \\
      Diffraction & 18.26 & 6.33  & -5.90  & -46.99 \\
      Unidentified & 20.80 & 37.33 & 34.54 & -43.81 \\
      Post B      & 21.23 & 23.75 & -29.37 & -44.03 \\
      Unidentified  & 18.79 & -3.98 & -27.59 & -49.67 \\
     \bottomrule
  \end{tabular}
\end{table}

\begin{table*}[!t]
  \centering
  \caption{Rimax estimates of the reflections from two poles in the two-pole experiment in an anechoic chamber, estimate/true}
  \label{Tab:two_pol_twoReflect}
  \begin{tabular}{c|cccc}
    \toprule
     Post A position & Post A $\varphi_T$ (deg) & Post A $\varphi_R$ (deg) & Post B $\varphi_T$ (deg) & Post B $\varphi_R$ (deg) \\
    \midrule
     1  & 40.10/45.89 & -17.44/-17.12 & 23.75/28.64 & -29.37/-27.57 \\
     2  & 37.11/42.44 & -20.38/-19.22 & 24.94/28.64 & -28.72/-27.57 \\
     3  & 34.48/39.53 & -22.81/-21.66 & 21.99/28.64 & -28.94/-27.57 \\
     4  & 30.41/36.50 & -27.05/-22.98 & NA/28.64 & NA/-27.57 \\
     5  & NA/31.35 & NA/-22.77 & 24.92/28.64 & -28.60/-27.57 \\
    \bottomrule
  \end{tabular}
\end{table*}

\section{Conclusion}
In this paper, we presented the calibration procedures and the verification experiments for a \gls{mimo} \gls{mmwave} channel sounder based on phased arrays, in order to demonstrate its capability to perform an \gls{hrpe} evaluation such as Rimax {\color{black}or \gls{sage}}. {\color{black}The integrated design of antenna array with other RF electronics requires us to rethink the traditional sounder calibration procedures.} As a result, we proposed a novel two-setup calibration scheme and formulated the extraction and separation of system frequency response and the beam pattern as two optimization problems. 

We also investigated two practical issues that might impact \gls{mmwave} array calibration, such as the array center misalignment and the phase noise variation. We also conclude through simulations that the fast-varying phase noise could lead to a limitation of the dynamic range to \SI{18}{dB}, a similar level that we observed in the chamber verification experiment. {\color{black}The two-path channel measurement in an anechoic chamber show that the sounder with the calibration results can perform \gls{hrpe} evaluation, and the estimates of strong paths match well with the environment. The two-pole channel measurement suggests that the calibration enables the sounder to differentiate signals from two reflectors which are only about $5^\circ$ apart in azimuth \gls{doa}.}

{\color{black}In the future, we will study if there is an optimal set of beams to estimate propagation paths. Besides we will investigate the optimal bandwidth in the narrowband pattern extraction algorithm. We will also investigate how to apply the wideband Rimax to the data evaluation and if the calibration procedures can therefore be simplified. }

\bibliographystyle{IEEEtran}
\bibliography{enHRPE_main}

\begin{thebibliography}{10}
\providecommand{\url}[1]{#1}
\csname url@samestyle\endcsname
\providecommand{\newblock}{\relax}
\providecommand{\bibinfo}[2]{#2}
\providecommand{\BIBentrySTDinterwordspacing}{\spaceskip=0pt\relax}
\providecommand{\BIBentryALTinterwordstretchfactor}{4}
\providecommand{\BIBentryALTinterwordspacing}{\spaceskip=\fontdimen2\font plus
\BIBentryALTinterwordstretchfactor\fontdimen3\font minus
  \fontdimen4\font\relax}
\providecommand{\BIBforeignlanguage}[2]{{%
\expandafter\ifx\csname l@#1\endcsname\relax
\typeout{** WARNING: IEEEtran.bst: No hyphenation pattern has been}%
\typeout{** loaded for the language `#1'. Using the pattern for}%
\typeout{** the default language instead.}%
\else
\language=\csname l@#1\endcsname
\fi
#2}}
\providecommand{\BIBdecl}{\relax}
\BIBdecl

\bibitem{andrews2014will}
J.~G. Andrews, S.~Buzzi, W.~Choi, S.~V. Hanly, A.~Lozano, A.~C. Soong, and
  J.~C. Zhang, ``What will 5g be?'' \emph{IEEE Journal on selected areas in
  communications}, vol.~32, no.~6, pp. 1065--1082, 2014.

\bibitem{rappaport2014millimeter}
T.~S. Rappaport, R.~W. Heath~Jr, R.~C. Daniels, and J.~N. Murdock,
  \emph{Millimeter wave wireless communications}.\hskip 1em plus 0.5em minus
  0.4em\relax Pearson Education, 2014.

\bibitem{Molisch_book_2000}
A.~F. Molisch, A.~Mammela, and D.~P. Taylor, \emph{Wideband wireless digital
  communication}.\hskip 1em plus 0.5em minus 0.4em\relax prentice hall PTR,
  2000.

\bibitem{samimi201328}
M.~Samimi, K.~Wang, Y.~Azar, G.~N. Wong, R.~Mayzus, H.~Zhao, J.~K. Schulz,
  S.~Sun, F.~Gutierrez, and T.~S. Rappaport, ``28 ghz angle of arrival and
  angle of departure analysis for outdoor cellular communications using
  steerable beam antennas in new york city,'' in \emph{Vehicular Technology
  Conference (VTC Spring), 2013 IEEE 77th}.\hskip 1em plus 0.5em minus
  0.4em\relax IEEE, 2013, pp. 1--6.

\bibitem{3gppAbove6GHz}
3GPP, ``Technical specification group radio access network; study on channel
  model for frequency spectrum above 6 ghz. tr 38.900,'' 3rd Generation
  Partnership Project (3GPP), Tech. Rep., July 2017.

\bibitem{metis2020}
METIS2020, ``Metis channel model,'' Tech. Rep. METIS2020, Tech. Rep.,
  Delivierable D1.4 v3, July 2015 [Online]. Available:
  https://www.metis2020.com/wp-content/uploads/METIS.

\bibitem{ic1004}
``http://www.ic1004.org/.''

\bibitem{NIST5G}
NIST, ``5g mmwave channel model,
  http://www.nist.gov/ctl/wireless-networks/5gmillimeterwavechannelmodel.cfm.''

\bibitem{haneda2015channel}
K.~Haneda, ``Channel models and beamforming at millimeter-wave frequency
  bands,'' \emph{IEICE Transactions on Communications}, vol.~98, no.~5, pp.
  755--772, 2015.

\bibitem{rappaport2015wideband}
T.~S. Rappaport, G.~R. MacCartney, M.~K. Samimi, and S.~Sun, ``Wideband
  millimeter-wave propagation measurements and channel models for future
  wireless communication system design,'' \emph{IEEE Transactions on
  Communications}, vol.~63, no.~9, pp. 3029--3056, 2015.

\bibitem{hur2014synchronous}
S.~Hur, Y.-J. Cho, J.~Lee, N.-G. Kang, J.~Park, and H.~Benn, ``Synchronous
  channel sounder using horn antenna and indoor measurements on 28 ghz,'' in
  \emph{Communications and Networking (BlackSeaCom), 2014 IEEE International
  Black Sea Conference on}.\hskip 1em plus 0.5em minus 0.4em\relax IEEE, 2014,
  pp. 83--87.

\bibitem{kim201528ghz}
M.-D. Kim, J.~Liang, Y.~K. Yoon, and J.~H. Kim, ``28ghz path loss measurements
  in urban environments using wideband channel sounder,'' in \emph{Antennas and
  Propagation \& USNC/URSI National Radio Science Meeting, 2015 IEEE
  International Symposium on}.\hskip 1em plus 0.5em minus 0.4em\relax IEEE,
  2015, pp. 1798--1799.

\bibitem{bas2017real}
C.~Bas, R.~Wang, D.~Psychoudakis, T.~Henige, R.~Monroe, J.~Park, J.~Zhang, and
  A.~Molisch, ``A real-time millimeter-wave phased array mimo channel
  sounder,'' in \emph{Vehicular Technology Conference (VTC-Fall), 2017 IEEE
  86th}.\hskip 1em plus 0.5em minus 0.4em\relax IEEE, 2017, pp. 1--6.

\bibitem{akdeniz2014millimeter}
M.~R. Akdeniz, Y.~Liu, M.~K. Samimi, S.~Sun, S.~Rangan, T.~S. Rappaport, and
  E.~Erkip, ``Millimeter wave channel modeling and cellular capacity
  evaluation,'' \emph{IEEE journal on selected areas in communications},
  vol.~32, no.~6, pp. 1164--1179, 2014.

\bibitem{papazian2016radio}
P.~B. Papazian, C.~Gentile, K.~A. Remley, J.~Senic, and N.~Golmie, ``A radio
  channel sounder for mobile millimeter-wave communications: System
  implementation and measurement assessment,'' \emph{IEEE Transactions on
  Microwave Theory and Techniques}, vol.~64, no.~9, pp. 2924--2932, 2016.

\bibitem{fleury2002high}
B.~H. Fleury, P.~Jourdan, and A.~Stucki, ``High-resolution channel parameter
  estimation for {MIMO} applications using the {SAGE} algorithm,'' in
  \emph{Broadband Communications, 2002. Access, Transmission, Networking. 2002
  International Zurich Seminar on}.\hskip 1em plus 0.5em minus 0.4em\relax
  IEEE, 2002, pp. 30--1.

\bibitem{gustafson2011directional}
C.~Gustafson, F.~Tufvesson, S.~Wyne, K.~Haneda, and A.~F. Molisch,
  ``Directional analysis of measured 60 ghz indoor radio channels using sage,''
  in \emph{Vehicular Technology Conference (VTC Spring), 2011 IEEE 73rd}.\hskip
  1em plus 0.5em minus 0.4em\relax IEEE, 2011, pp. 1--5.

\bibitem{gustafson2014mm}
C.~Gustafson, K.~Haneda, S.~Wyne, and F.~Tufvesson, ``On mm-wave multipath
  clustering and channel modeling,'' \emph{IEEE Transactions on Antennas and
  Propagation}, vol.~62, no.~3, pp. 1445--1455, 2014.

\bibitem{Yin_et_al_EuCAP2014}
X.~Yin, Y.~He, Z.~Song, M.~D. Kim, and H.~K. Chung, ``A
  sliding-correlator-based sage algorithm for mm-wave wideband channel
  parameter estimation,'' in \emph{The 8th European Conference on Antennas and
  Propagation (EuCAP 2014)}, April 2014, pp. 625--629.

\bibitem{martinez2014deterministic}
M.-T. Martinez-Ingles, D.~P. Gaillot, J.~Pascual-Garcia, J.-M.
  Molina-Garcia-Pardo, M.~Lienard, and J.-V. Rodr{\'\i}guez, ``Deterministic
  and experimental indoor mmw channel modeling,'' \emph{IEEE Antennas and
  Wireless Propagation Letters}, vol.~13, pp. 1047--1050, 2014.

\bibitem{psychoudakis2016mobile}
D.~Psychoudakis, H.~Zhou, B.~Biglarbegian, T.~Henige, and F.~Aryanfar, ``Mobile
  station radio frequency unit for 5g communications at 28ghz,'' in
  \emph{Microwave Symposium (IMS), 2016 IEEE MTT-S International}.\hskip 1em
  plus 0.5em minus 0.4em\relax IEEE, 2016, pp. 1--3.

\bibitem{papazian2015radio}
P.~B. Papazian, K.~A. Remley, C.~Gentile, and N.~Golmie, ``Radio channel
  sounders for modeling mobile communications at 28 ghz, 60 ghz and 83 ghz,''
  in \emph{Millimeter Waves (GSMM), 2015 Global Symposium On}.\hskip 1em plus
  0.5em minus 0.4em\relax IEEE, 2015, pp. 1--3.

\bibitem{Steinbauer_et_al_2001}
M.~Steinbauer, A.~F. Molisch, and E.~Bonek, ``The double-directional radio
  channel,'' \emph{IEEE Antennas and propagation Magazine}, vol.~43, no.~4, pp.
  51--63, 2001.

\bibitem{ji2017antenna}
Y.~Ji, X.~Yin, H.~Wang, S.~X. Lu, and C.~Cao, ``Antenna de-embedded
  characterization for 13--17-ghz wave propagation in indoor environments,''
  \emph{IEEE Antennas and Wireless Propagation Letters}, vol.~16, pp. 42--45,
  2017.

\bibitem{maccartney2017flexible}
G.~R. MacCartney and T.~S. Rappaport, ``A flexible millimeter-wave channel
  sounder with absolute timing,'' \emph{IEEE Journal on Selected Areas in
  Communications}, vol.~35, no.~6, pp. 1402--1418, 2017.

\bibitem{wen2016mmwave}
Z.~Wen, H.~Kong, Q.~Wang, S.~Li, X.~Zhao, M.~Wang, and S.~Sun, ``mmwave channel
  sounder based on cots instruments for 5g and indoor channel measurement,'' in
  \emph{Wireless Communications and Networking Conference Workshops (WCNCW),
  2016 IEEE}.\hskip 1em plus 0.5em minus 0.4em\relax IEEE, 2016, pp. 37--43.

\bibitem{papazian2016calibration}
P.~B. Papazian, J.-K. Choi, J.~Senic, P.~Jeavons, C.~Gentile, N.~Golmie,
  R.~Sun, D.~Novotny, and K.~A. Remley, ``Calibration of millimeter-wave
  channel sounders for super-resolution multipath component extraction,'' in
  \emph{Antennas and Propagation (EuCAP), 2016 10th European Conference
  on}.\hskip 1em plus 0.5em minus 0.4em\relax IEEE, 2016, pp. 1--5.

\bibitem{sayeed2002deconstructing}
A.~M. Sayeed, ``Deconstructing multiantenna fading channels,'' \emph{IEEE
  Transactions on Signal Processing}, vol.~50, no.~10, pp. 2563--2579, 2002.

\bibitem{richter2005estimation}
A.~Richter, ``Estimation of radio channel parameters: Models and algorithms,''
  Ph.D. dissertation, Techn. Univ. Ilmenau, Ilmenau, Germany, May 2005.
  [Online]. Available: http://www.db-thueringen.de.

\bibitem{salmi2009detection}
J.~Salmi, A.~Richter, and V.~Koivunen, ``Detection and tracking of {MIMO}
  propagation path parameters using state-space approach,'' \emph{Signal
  Processing, IEEE Transactions on}, vol.~57, no.~4, pp. 1538--1550, 2009.

\bibitem{richter2005joint}
A.~Richter and R.~S. Thoma, ``Joint maximum likelihood estimation of specular
  paths and distributed diffuse scattering,'' in \emph{Vehicular Technology
  Conference, 2005. VTC 2005-Spring. 2005 IEEE 61st}, vol.~1.\hskip 1em plus
  0.5em minus 0.4em\relax IEEE, 2005, pp. 11--15.

\bibitem{salmi2011propagation}
J.~Salmi and A.~F. Molisch, ``Propagation parameter estimation, modeling and
  measurements for ultrawideband mimo radar,'' \emph{IEEE Transactions on
  Antennas and Propagation}, vol.~59, no.~11, pp. 4257--4267, 2011.

\bibitem{landmann2012impact}
M.~Landmann, M.~Kaske, and R.~S. Thoma, ``Impact of incomplete and inaccurate
  data models on high resolution parameter estimation in multidimensional
  channel sounding,'' \emph{IEEE Transactions on Antennas and Propagation},
  vol.~60, no.~2, pp. 557--573, 2012.

\bibitem{belloni2007doa}
F.~Belloni, A.~Richter, and V.~Koivunen, ``Doa estimation via manifold
  separation for arbitrary array structures,'' \emph{IEEE Transactions on
  Signal Processing}, vol.~55, no.~10, pp. 4800--4810, 2007.

\bibitem{eckart1936approximation}
C.~Eckart and G.~Young, ``The approximation of one matrix by another of lower
  rank,'' \emph{Psychometrika}, vol.~1, no.~3, pp. 211--218, 1936.

\bibitem{dunsmore2011new}
J.~Dunsmore, ``A new calibration method for mixer delay measurements that
  requires no calibration mixer,'' in \emph{Microwave Conference (EuMC), 2011
  41st European}.\hskip 1em plus 0.5em minus 0.4em\relax IEEE, 2011, pp.
  480--483.

\bibitem{KT-FCAaccuracy}
``Imporving measurement and calibration accuracy using the frequency converter
  application in the pna microwave network analyzer,'' Keysight Technologies,
  Tech. Rep., Application Note.

\bibitem{bas2018real_arxiv}
C.~U. Bas, R.~Wang, S.~Sangodoyin, D.~Psychoudakis, T.~Henige, R.~Monroe,
  J.~Park, J.~Zhang, and A.~F. Molisch, ``Real-time millimeter-wave mimo
  channel sounder for dynamic directional measurements,'' \emph{arXiv preprint
  arXiv:1807.11921}, 2018.

\bibitem{marquardt1963algorithm}
D.~W. Marquardt, ``An algorithm for least-squares estimation of nonlinear
  parameters,'' \emph{Journal of the society for Industrial and Applied
  Mathematics}, vol.~11, no.~2, pp. 431--441, 1963.

\bibitem{prata2002misaligned}
A.~Prata~Jr, ``Misaligned antenna phase-center determination using measured
  phase patterns,'' \emph{Interplanetary Network Progress Report}, vol. 150,
  no.~27, pp. 1--9, 2002.

\bibitem{chen2014determining}
Y.~Chen and R.~G. Vaughan, ``Determining the three-dimensional phase center of
  an antenna,'' in \emph{General Assembly and Scientific Symposium (URSI GASS),
  2014 XXXIth URSI}.\hskip 1em plus 0.5em minus 0.4em\relax IEEE, 2014, pp.
  1--4.

\bibitem{strang1993introduction}
G.~Strang, G.~Strang, G.~Strang, and G.~Strang, \emph{Introduction to linear
  algebra}.\hskip 1em plus 0.5em minus 0.4em\relax Wellesley-Cambridge Press
  Wellesley, MA, 1993, vol.~3.

\bibitem{landmann2006estimation}
M.~Landmann and R.~Thom{\"a}, ``Estimation of phase drift during calibration
  measurements for efficient beam pattern modelling,'' in \emph{Proceedings of
  NEWCOM-ACoRN Workshop}, 2006.

\bibitem{wang2018channel}
R.~Wang, O.~Renaudin, C.~U. Bas, S.~Sangodoyin, and A.~F. Molisch, ``On channel
  sounding with switched arrays in fast time-varying channels,'' \emph{arXiv
  preprint arXiv:1805.08069}, 2018.

\bibitem{wang2015efficiency}
R.~Wang, O.~Renaudin, R.~M. Bernas, and A.~F. Molisch, ``Efficiency improvement
  for path detection and tracking algorithm in a time-varying channel,'' in
  \emph{Vehicular Technology Conference (VTC Fall), 2015 IEEE 82nd}.\hskip 1em
  plus 0.5em minus 0.4em\relax IEEE, 2015, pp. 1--5.

\bibitem{richter2006distributed}
A.~Richter, J.~Salmi, and V.~Koivunen, ``Distributed scattering in radio
  channels and its contribution to mimo channel capacity,'' in \emph{Antennas
  and Propagation, 2006. EuCAP 2006. First European Conference on}.\hskip 1em
  plus 0.5em minus 0.4em\relax IEEE, 2006, pp. 1--7.

\bibitem{poutanen2011angular}
J.~Poutanen, J.~Salmi, K.~Haneda, V.-M. Kolmonen, and P.~Vainikainen, ``Angular
  and shadowing characteristics of dense multipath components in indoor radio
  channels,'' \emph{IEEE Transactions on Antennas and Propagation}, vol.~59,
  no.~1, pp. 245--253, 2011.

\end{thebibliography}

\end{document}